\documentclass[article,11pt]{article}

\newcommand{\longversion}[1]{#1}
\newcommand{\shortversion}[1]{}

    \usepackage[preprint,nonatbib]{neurips_2022}

    \usepackage[sort,numbers]{natbib}

\usepackage[utf8]{inputenc} 
\usepackage[T1]{fontenc}    
\usepackage{hyperref}       
\usepackage{url}            
\usepackage{booktabs}       
\usepackage{amsfonts}       
\usepackage{nicefrac}       
\usepackage{microtype}      
\usepackage{xcolor}         
\usepackage{lineno}

\usepackage[normalem]{ulem}
\usepackage{color}
\definecolor{blue(ncs)}{rgb}{0.0, 0.53, 0.74}
\usepackage{amssymb}
\usepackage{amsmath}
\usepackage[linesnumbered,ruled,vlined]{algorithm2e}
\DontPrintSemicolon
\SetKwInOut{Parameter}{Parameter}
\usepackage[belowskip=-8pt]{caption}
\usepackage{wrapfig}
\usepackage{subcaption}
\usepackage{colortbl}
\definecolor{webgreen}{rgb}{0,0.4,0}
\definecolor{webbrown}{rgb}{0.6,0,0}
\definecolor{purple}{rgb}{0.5,0,0.25}
\definecolor{darkblue}{rgb}{0,0,0.7}
\definecolor{darkred}{rgb}{0.7,0,0}
\usepackage{hyperref}
\hypersetup{colorlinks,citecolor=darkblue,filecolor=black,linkcolor=darkred,urlcolor=webgreen,pdfpagemode=UseNone, pdfstartview=FitH}
\usepackage[textsize=small]{todonotes}

\newcommand{\ignore}[1]{}

\newtheorem{lemma}{{Lemma}}
\newtheorem{corollary}{{Corollary}}
\newtheorem{theorem}{{Theorem}}
\newtheorem{definition}{{Definition}}

\newtheorem{fact}{{Fact}}

\usepackage{cleveref}
\crefname{claim}{claim}{claims}
\crefname{fact}{fact}{facts}
\crefname{algorithm}{alg.}{algs.}
\crefname{observation}{observation}{observations}
\crefname{equation}{eqn.}{eqns.}
\crefname{assumption}{assumption}{assumptions}
\crefname{lemma}{lemma}{lemmata}
\crefname{corollary}{corollary}{corollaries}
\crefname{figure}{fig.}{figs.}
\crefname{definition}{def.}{defs.}

\newenvironment{proof}{\noindent {\em Proof\/}:\enspace}
{\hfill $\blacksquare{}$}
\newenvironment{psketch}{\noindent {\em Proof sketch\/}:\enspace}
{\hfill $\blacksquare{}$}

\DeclareMathOperator*{\argmin}{\arg\!\min}
\DeclareMathOperator*{\argmax}{\arg\!\max}

\usepackage{pgf}
\usepackage{verbatim}
\usepackage[shortlabels]{enumitem}
\usepackage{amssymb}
\usepackage{mathrsfs}
\usepackage{mathtools}

\usepackage{resizegather}

\newif\ifverbose
\verbosetrue 
\newcommand{\sn}[1]  {\ifverbose {\noindent \textcolor{blue}{{\bf SN: }\emph{#1}} } \else \fi }

\newcommand{\edit}[1]  {\ifverbose{{\color{red}#1}\color{black}}\else {{#1}}\fi}

\usepackage{url}

\usepackage[cal=esstix,frak=euler,scr=boondox,bb=pazo]{mathalfa}
\usepackage{xspace}

\newcommand{\gggg}{\ensuremath{\mathscr{g}}\xspace}
\newcommand{\pppp}{\ensuremath{\mathscr{p}}\xspace}

\usepackage{nicefrac}

\verbosefalse 

\usepackage{xcolor, soul}
\sethlcolor{blue}
\usepackage{cleveref}
\crefname{claim}{claim}{claims}
\crefname{fact}{fact}{facts}
\crefname{algorithm}{algorithm}{algorithms}
\crefname{observation}{observation}{observations}
\crefname{equation}{equation}{equations}
\crefname{assumption}{assumption}{assumptions}
\crefname{hypothesis}{hypothesis}{hypotheses}
\usepackage{subcaption}
\usepackage[noend]{algorithmic}
\usepackage{resizegather}
\usepackage{enumitem}
\setlist{topsep=0pt, leftmargin=*}
\usepackage{multirow,bigdelim}
\usepackage{tikz}
\usepackage{framed}
\usepackage{wrapfig}
\usepackage{cutwin}

\usepackage[compact]{titlesec}
\titlespacing{\section}{2pt}{2pt}{2pt}
\titlespacing{\subsection}{2pt}{2pt}{2pt}

\usepackage{tfrupee}
\usepackage{multirow}
\usepackage{authblk}
\usepackage{multicol}

\renewenvironment{itemize}{
  \begin{list}{\textbullet}{
    \setlength{\leftmargin}{1em}
    \setlength{\itemsep}{0.25em}
    \setlength{\parskip}{0pt}
    \setlength{\parsep}{0pt}
  }
}{
  \end{list}
}

\newcommand{\da}{\text{\bf \texttt{DA}}}
\newcommand{\ta}{\text{\bf \texttt{TA}}}
\newcommand{\ddsic}{\textup{\bf \texttt{DDSIC}}}
\newcommand{\dbic}{\textup{\bf \texttt{DBIC}}}
\newcommand{\lev}{\textup{\bf \texttt{LbLEV}}}
\newcommand{\levg}{\textup{\bf \texttt{LbL-Gen}}}
\newcommand{\ra}{\textup{\bf \texttt{RA}}}
\newcommand{\idm}{\textup{\bf \texttt{IDM}}}
\newcommand{\tnm}{\textup{\bf \texttt{TNM}}}
\newcommand{\vipc}{\textup{\bf \texttt{VIPC}}}
\newcommand{\ppp}{\textup{\bf \texttt{pay}}}
\newcommand{\mvv}{\textup{\bf \texttt{maxViVa}}}
\newcommand{\of}{\textup{\bf \texttt{offset}}}
\newcommand{\pa}{\textup{\bf \texttt{parent}}}

\newcommand{\mff}{\textup{\bf \texttt{MFF}}}
\newcommand{\mffe}{\textup{\bf \texttt{MFFE}}}

\newcommand{\ch}{\textup{\bf \texttt{children}}}
\newcommand{\wi}{\textup{\bf \texttt{winner}}}
\newcommand{\ru}{\textup{\bf \texttt{runnerup}}}
\newcommand{\win}{\emph{winner}}
\newcommand{\opnw}{\emph{on-path non-winner}}
\newcommand{\nopnw}{\emph{not-on-path non-winner}}
\newcommand{\Nfirst}{{\tilde{N}}}
\newcommand*\diff{\mathop{}\!\mathrm{d}}

\allowdisplaybreaks

\title{\bf \edit{Optimal Referral Auction Design}}

\author[1]{Rangeet Bhattacharya}
\author[2]{Parvik Dave}
\author[3]{Palash Dey}
\author[2]{Swaprava Nath}

\affil[1]{\small Indian Institute of Technology Kanpur, \texttt{rangebhat@cse.iitk.ac.in} }
\affil[2]{\small Indian Institute of Technology Bombay, \texttt{\{parvikdave,swaprava\}@iitb.ac.in} }
\affil[3]{\small Indian Institute of Technology Kharagpur, \texttt{palash.dey@cse.iitkgp.ac.in} }
\date{\today}

\begin{document}
\maketitle              
\begin{abstract}
The auction of a single indivisible item is one of the most celebrated problems in mechanism design with transfers. Despite its simplicity, it provides arguably the cleanest and most insightful results in the literature. When the information that {\em the auction is running} is available to every participant, \citet{Myerson1981} provided a seminal result to characterize the incentive-compatible auctions along with revenue optimality. However, such a result does not hold in an {\em auction on a network}, where the information of the auction is spread via the agents, and they need incentives to forward the information. In recent times, a few auctions (e.g., \cite{li2017mechanism,li2022diffusion}) were designed that appropriately incentivized the intermediate nodes on the network to {\em promulgate} the information to potentially more valuable bidders. In this paper, we provide a Myerson-like characterization of {\em incentive-compatible auctions on a network} and show that the currently known auctions fall within this class of randomized auctions. 
\edit{We then consider a special class called the {\em referral auctions} that are inspired by the multi-level marketing mechanisms~\citep{Emek11,Drucker2012,Babaioff2012} and obtain the structure of a revenue optimal referral auction for i.i.d.\ bidders.
Through experiments, we show that even for non-i.i.d.\ bidders there exist auctions following this characterization that can provide a higher revenue than the currently known auctions on networks.}
\end{abstract}


\sn{modified presentation plan:
\begin{itemize}
    \item characterization of truthful diffusion auctions.
    \item non-trivial truthful diffusion auction that yields more revenue than IDM for non-i.i.d.\ bidders on trees (empirical verification).
    \item consider multi-level auctions (MLA), inspired by multi-level mechanisms, find the optimal revenue MLA for i.i.d.\ bidders.
    \item the optimal mechanism turns out to be a class of mechanisms that admits IDM with a reserve price.
    \item interesting to note that IDM fetches the worst revenue in CDM class, but the reserve price variant is the optimal MLA.
    \item in the shorter version, we can remove the \lev{} altogether and give the randomized \ddsic{} example, complete details of the revenue optimality in the class of MLA. providing proofs if needed.
\end{itemize}
}

%
%
%
\section{Introduction}
\label{sec:intro}

Single indivisible item auction is a special setting of mechanism design with monetary transfers where multiple bidders contest to collect a single item. The {\em true value} of the item could be different for different agents and it is their {\em private information}, i.e., not known to the designer of a {\em mechanism}.\footnote{Since auctions are special cases of mechanisms, we will use these two terms interchangeably in this paper.}
Despite its simplicity, single-item auction provides remarkable insights into the following questions: (a)~what is the structure of the mechanisms that reveal the agents' true private information, (b)~how to design mechanisms that maximize the expected revenue. In a world where the information that `an item is being auctioned' is available to every possible bidder interested in this item, these two questions have been answered gracefully by Myerson in his seminal paper~\citep{Myerson1981}.

However, in various recent contexts of auctions, the network of connections makes an important role in the information flow over the network. An agent diffuses the information into the network only if it finds it is {\em beneficial} to share. This setup is called {\em network auctions}, where agents {\em diffuse} the information of the auction only if it (strictly or weakly) improves their utilities. This problem has given birth to the domain of {\em diffusion auction design} on networks and has received significant attention in the recent times~\citep{li2017mechanism,yuhangguodonghao2021,li2022diffusion,li2020incentive}. Because the information about the auction does not {\em automatically} reach every agent in this setup, the mechanism needs to incentivize the individuals to diffuse (or forward) the information. The \citet{Myerson1981} characterization does not follow here, and a fresh investigation is necessary to characterize the {\em truthful} and {\em revenue maximizing} diffusion auctions.

\subsection{Our contributions}
\label{sec:contributions}

The contributions of this paper are divided into two parts. In the first part, we characterize the {\em truthful} network auction via certain constraints on allocation and payment. In the second part, we consider a Bayesian setup and find the class of revenue-optimal network auctions.
More concretely:
\begin{enumerate}
    \item We provide a more direct definition of truthfulness called {\em diffusion dominant strategy incentive compatibility} (\ddsic, \Cref{def:DDSIC}) and show that it is equivalent to the existing IC definition in the literature on network auction~\cite[e.g.]{li2020incentive}.
    \item We characterize \ddsic{} (and therefore, IC) mechanisms (\Cref{thm:characterization,thm:reverse_characterization,thm:ddsic-ic,cor:equivalence}) with constraints on the allocation and payment of the auctions. \edit{Note that this result is of independent interest irrespective of the revenue optimality question addressed later in this paper (similar to \citep{Myerson1981}).}
    \item We introduce a new mechanism called \lev\ that is \ddsic\ (\Cref{thm:lblev-truthful}) and individually rational (\Cref{thm:lblev-ir}) on a tree.
    \item \edit{We find the revenue-optimal {\em referral auction} (which is a class of auctions motivated by the multi-level marketing methods) for i.i.d.\ bidders (\Cref{thm:optimal-revenue-tree}) in \S\ref{sec:optimal-trees}.}
    \item When bidders are non-i.i.d., we experimentally exhibit that the parameters of \lev\ can be tuned, based on the prior information of the valuations and network structure, to yield a better revenue than the currently known truthful diffusion auctions (\S\ref{sec:experiment}).
\end{enumerate}
\shortversion{For the dearth of space, we have moved the detailed proofs of some of the theoretical results to the supplementary material.}

\subsection{Related work}
\label{sec:literature}

The area of diffusion auction design is relatively new, leading to a rather thin literature. \citet{yuhangguodonghao2021} provide a comprehensive survey of the domain. The first works on diffusion auction are due to \citet{li2017mechanism} and \citet{lee2016mechanisms}. In particular, \cite{li2017mechanism} showed that the classical VCG mechanism \citep{Vick61,Clar71,Grov73} can be extended to the diffusion setting, but it may lead to a large deficit. They propose a new mechanism called IDM that mitigates this problem. In the following years, a few more diffusion auctions were proposed: CSM~\citep{li2018customer} for economic networks, MLDM for intermediary networks~\citep{li2020information}, TNM, CDM, WDM were on the unweighted and weighted networks~\citep{li2019diffusion,li2022diffusion}, FDM~\citep{zhang2020redistribution} and NRM~\citep{zhang2020incentivize} considered the money burning issue in network auction and proposed schemes to redistribute the money maintaining incentive compatibility. On the characterization results, \citet{li2020incentive} provide a characterization for deterministic diffusion auctions and find optimal payments. Our approach, however, considers a broader approach to characterize all {\em randomized} diffusion auctions (which includes the deterministic auctions as a special case) and shows that better revenue-generating mechanisms can be found.

\edit{Diffusion auction has a close similarity with a business method called {\em multi-level marketing} (MLM) (a survey can be found in \citep{reingewertz2021economic}). Direct sales firms often use this method to encourage individual distributors to recruit new distributors. It is a multi-billion dollar industry (US figures available in~\citep{nat2002marketing}) -- according to a 2018 survey, 7.7\% of the US adult population had participated in at least one MLM business during their lifetime~\citep{deliema2018aarp}.
A prominent example of an MLM scheme is the DARPA red balloon challenge.\footnote{\url{https://www.darpa.mil/about-us/timeline/network-challenge}}
MLM mechanisms are also well investigated in the mechanism design literature~\citep{Emek11,Drucker2012,Babaioff2012}. Despite their criticism for being similar to pyramid schemes, \citet{nat2002marketing} show some important differences that keep MLM mechanisms relevant in practice. A diffusion auction incentivizes agents on a network to refer to more individuals who are potentially interested in participating in the auction. Hence, in this paper, we consider a natural candidate class of {\em referral auctions} for revenue optimality~(\S~\ref{sec:referral}).
}

\section{Basic Problem Setup}
\label{sec:setup}

Consider a directed graph $G = (N \cup \{s\}, E)$, where $N = \{1,\ldots,n\}$ is the set of players involved in the auction of a single indivisible item and $s$ is a distinguished node called the {\em seller}. The set $E$ is the set of edges. Each edge $(i,j)$ denotes that if node $i$ shares information, and node $j$ will receive it. Typical examples of such graphs are online social networks where an individual can share information (selectively) with a subset of her neighbors. The direction signifies that almost all networks have asymmetric information flow (e.g., only followers receive the information from the followee).

In this network, node $s$ is a single seller that wants to sell the indivisible item. Every other node $i \in N$ is a potential buyer and the information about the auction flows only via the direction of an edge. The information cannot reach a node unless there is a directed path from $s$ to that node and each intermediate node decides to forward the information. An intermediate node may decide not to forward the information if it reduces its {\em utility}.

This setup naturally brings up an auction-like information-sharing game among the players. Each player $i \in N$ has a type $\theta_i = (v_i, r_i)$, where $v_i$ is the valuation of agent $i$ for the item, and $r_i$ is the set of her directed neighbors. The set of  valuations and subsets of neighbors of player $i$ are denoted by ${\cal V}_i$ and ${\cal R}_i$ respectively. The type set of $i$, $\Theta_i$, is therefore, ${\cal V}_i \times {\cal R}_i$, and agent $i$ can report its type from this set. The information about the auction needs to reach via directed edges to player $i$ for her to participate in the auction. Therefore, the auction asks every agent to report her valuation for the item and to forward the information to its directed neighbors. In our model, this is captured via their reported type $\hat{\theta}_i = (\hat{v}_i, \hat{r}_i)$ for every agent $i \in N$. We assume that the seller $s$ is not a strategic player in this auction, rather he wants to sell the object and always forwards the information to his directed neighbors. The vector of the reported types of all the agents except $i$ is denoted by $\hat{\theta}_{-i} = (\hat{\theta}_{1}, \ldots, \hat{\theta}_{i-1}, \hat{\theta}_{i+1}, \ldots, \hat{\theta}_{n})$. We denote the set of all type profiles by $\Theta := \prod_{i \in N} \Theta_i$.

Depending on the reported types of the agents, particularly, the reported $\hat{r}_i$'s, the auction may reach only a subset of the agents in $N$. Throughout this paper, we will use the notation $r_i$ to denote the {\em true} neighbor set of player $i$ and assume that the reported $\hat{r}_i$ can only be a subset of it. To denote the reported valuation and directed neighbors on the subnetwork generated by $(\hat{\theta}_{i}, \hat{\theta}_{-i})$, we use a {\em filter function} $f^G$ for the graph $G$, where $f^G(\hat{\theta}_{i}, \hat{\theta}_{-i})$ denotes the reported valuation and directed neighbor vector of the subgraph where each node has a directed path from $s$ after the agents reported the type profile $(\hat{\theta}_{i}, \hat{\theta}_{-i})$.
%
In this setup, the auction design goal is to incentivize each node to truthfully reveal its private valuation and forward regardless of others' actions. It is known that such mechanism exists~ \citep{li2017mechanism}. One of the goals of this paper is to characterize all such mechanisms in an elegant manner.

We consider auctions on this graph with randomized allocations to the agents. Formally, we define a {\em diffusion auction} in this setup as follows.

\begin{definition}[Diffusion Auction]
\label{def:diff-auction}
 A {\em diffusion auction (\da)} is given by the tuple $(g, p)$ where $g$ and $p$ are the {\em allocation} and {\em payment} functions respectively. The allocation function $g : \Theta \to \Delta_n$ is such that its $i$-th component $g_i(f^G(\theta))$ denotes the probability of agent $i$ winning the object, where $\Delta_n := \{x \in \mathbb{R}_{\geqslant 0}^n : \sum_{i=1}^n x_i = 1\}$. Similarly, the payment function $p = (p_i)_{i \in N}$ is such that its $i$-th component $p_i : \Theta \to \mathbb{R}$ denotes the payment assigned to agent $i$.
\end{definition}
Note that $g_i$ should operate on the subnetwork that remains connected to $s$ after the agents choose their actions $\hat{\theta}$. Hence the notation $g_i(f^G (\cdot) )$ is used in the definition above. It is worth noting that the notation generalizes the one used by \citet{li2017mechanism}. The action chosen by player $i$ may change the actions available to the other players and it is succinctly captured by the filter function which also subsumes the definition in that paper.
Also, note that \da\ is different from the classical auction, because the types of each agent now contain both the valuation ($v_i$) and the information on forwarding ($r_i$).
The utility of agent $i$ under \da\ is given by the standard quasi-linear model~\citep{SL08}: $u_i^{(g,p)}((\hat{\theta}_{i}, \hat{\theta}_{-i}); \theta_i) = v_i g_i(f^G(\hat{\theta}_{i}, \hat{\theta}_{-i})) - p_i(f^G(\hat{\theta}_{i}, \hat{\theta}_{-i}))$.

\section{Design Desiderata}
\label{sec:desiderata}

The first desirable property of an auction is truthfulness. However, in the context of auctions on the network, we need to ensure that the mechanism also incentivizes the agents to forward the information in addition to being truthful about their valuations. The following definition captures both these aspects.


\begin{definition}[Diffusion Dominant Strategy Incentive Compatibility]
\label{def:DDSIC}
 A \da\ $(g,p)$ on a graph $G$ is {\em diffusion dominant strategy incentive compatible (\ddsic)} if
 \begin{enumerate}
     \item \label{point1}
     every agent's utility is maximized by reporting her true valuation irrespective of the diffusing status of herself and the other agents, i.e., for every $i \in N$, $\forall r_i, \hat{\theta}_{-i}$, the following holds
     \begin{align*}
     \lefteqn{
     v_i g_i(f^G((v_i,r_i'), \hat{\theta}_{-i})) - p_i(f^G((v_i,r_i'), \hat{\theta}_{-i}))  } \\
     &\geqslant v_i g_i(f^G((v_i',r_i'), \hat{\theta}_{-i})) - p_i(f^G((v_i',r_i'), \hat{\theta}_{-i})), \forall v_i, v_i', \hat{\theta}_{-i}, r_i' \subseteq r_i, \textup{ and,}
     \end{align*}
     \item \label{point2}
     for every true valuation, every agent's utility is maximized by diffusing to all its neighbors irrespective of the diffusion status of the other agents, i.e., for every $i \in N$, $\forall r_i, \hat{\theta}_{-i}$, the following holds
     \begin{align*}
         \lefteqn{ v_i g_i(f^G((v_i,r_i), \hat{\theta}_{-i})) - p_i(f^G((v_i,r_i), \hat{\theta}_{-i}))  } \nonumber \\
         &\geqslant v_i g_i(f^G((v_i,r_i'), \hat{\theta}_{-i})) - p_i(f^G((v_i,r_i'), \hat{\theta}_{-i})), \forall v_i, \hat{\theta}_{-i}, r_i' \subseteq r_i.
     \end{align*}
 \end{enumerate}
\end{definition}

We show later in this paper that the above definition is equivalent to the following definition of incentive compatibility (restated below with the notation of this paper) given by \citet{li2020incentive} and hence can be used interchangeably.

\begin{definition}[Incentive Compatibility \citep{li2020incentive}]
\label{def:IC-zhao}
 A \da\ $(g,p)$ on a graph $G$ is incentive-compatible (IC) if for every $i \in N$, $\forall r_i, \hat{\theta}_{-i}$, $v_i g_i(f^G((v_i,r_i), \hat{\theta}_{-i})) - p_i(f^G((v_i,r_i), \hat{\theta}_{-i})) \geqslant v_i g_i(f^G((v_i',r_i'), \hat{\theta}_{-i})) - p_i(f^G((v_i',r_i'), \hat{\theta}_{-i})), \forall v_i, v_i', \forall r_i' \subseteq r_i, \forall i \in N$.
\end{definition}

\paragraph{Why \ddsic{}?}
A natural question can arise: why do we introduce a new definition of truthfulness when there is an existing one, given that both are equivalent? This is because the new definition provides a more direct and intuitive way to understand the truthful reporting of valuation and diffusion. \ddsic{} does this by splitting the IC condition into two sets of inequalities as given in \Cref{def:DDSIC}. In our proofs, this definition makes the analysis of truthful mechanisms simpler. We will show that IC and \ddsic{} are equivalent and both are equivalent to the two conditions stated in \Cref{thm:characterization}, and will subject all our further analyses only to \ddsic{}.

The next desirable property deals with the participation guarantee of the agents.
\begin{definition}[Individual Rationality]
\label{def:IR}
 A \da\ $(g,p)$ on a graph $G$ is {\em individually rational (IR)} if  $v_i g_i(f^G((v_i,r_i), \hat{\theta}_{-i})) - p_i(f^G((v_i,r_i), \hat{\theta}_{-i})) \geqslant 0, \ \forall v_i, r_i, \hat{\theta}_{-i}, \forall i \in N$.
\end{definition}

\section{Characterization Results}
\label{sec:results}

Our first result is to characterize the IC diffusion auctions and show equivalence between IC and \ddsic{}.
The result by \citet{Myerson1981} in the single indivisible item auction setup implicitly assumes that {\em the knowledge of auction reaches all the players for free}, and therefore, no additional incentive is required for the agents to {\em diffuse} the information of the auction into the network. But in our setting, the information reaches an agent on a network only if {\em every} predecessor in {\em at least} one {\em path} from the seller to that agent forwards this information. Our result, therefore, generalizes Myerson's characterization result in network auctions.
For a cleaner presentation, we define the following class of payments.

\begin{definition}[Monotone and Forwarding-Friendliness (MFF)]
 \label{def:ffm}
 For a given network $G$, a \da\ $(g,p)$ is {\em monotone and forwarding-friendly (\mff)} if
 \begin{enumerate}[(a)]
    \item \label{cond1} the functions $g_i(f^G((v_i, r_i), \hat{\theta}_{-i}))$ are monotone non-decreasing in $v_i$, for all $r_i, \hat{\theta}_{-i}$, and $i \in N$, and for the given allocation function $g$, the payment $p_i$ for each player $i \in N$ is such that, for every $v_i, r_i,$ and $\hat{\theta}_{-i}$, the following two conditions hold.
     \item \label{ffm:b} For every $r_i' \subseteq r_i$, the following payment formula is satisfied.
\begin{gather}
    p_i(f^G((v_i,r_i'),\hat{\theta}_{-i}) ) = p_i(f^G((0,r_i'),\hat{\theta}_{-i})) + v_i g_i(f^G((v_i,r_i'),\hat{\theta}_{-i}))-\int_0^{v_i} g_i(f^G((y,r_i'),\hat{\theta}_{-i})) \diff y \label{eq:int-formula}
\end{gather}
     \item \label{ffm:a} For every $r_i' \subseteq r_i$, the values of $p_i(f^G((0,r_i'),\hat{\theta}_{-i}))$ and $p_i(f^G((0,r_i),\hat{\theta}_{-i}))$ satisfies the following inequality.
\begin{gather}
p_i(f^G((0,r_i'),\hat{\theta}_{-i})) - p_i(f^G((0,r_i),\hat{\theta}_{-i})) \geqslant \int_0^{v_i}  \left ( g_i(f^G((y,r_i'),\hat{\theta}_{-i})) - g_i(f^G((y,r_i),\hat{\theta}_{-i})) \right ) \diff y.\label{eq:constraint}
\end{gather}
 \end{enumerate}
\end{definition}


%

We will refer to $p_i(f^G((0,r_i),\hat{\theta}_{-i}))$, the first term on the RHS of \Cref{eq:int-formula}, as the {\em value independent payment component} (\vipc) in the rest of the paper, since this component of player $i$ is not dependent on the valuation of $i$.
\begin{theorem}[IC $\Rightarrow$ \mff{}]
\label{thm:characterization}
If a  \da\ $(g,p)$ is IC, then it is \mff{}.
\end{theorem}

\paragraph{Discussions}

The characterization result is much in the spirit of the result of \citet{Myerson1981}. However, there are the following important observations on these results.
\begin{itemize}
    \item The payment component of \mff{} (\Cref{def:ffm}) has two conditions given by \Cref{eq:constraint,eq:int-formula}. While \Cref{eq:int-formula} is reminiscent of \citet{Myerson1981}, the important difference here is in the \vipc\ terms.
    These terms in the payment formula of \citet{Myerson1981} were unrestricted for the characterization of DSIC. However, for \ddsic, \Cref{eq:constraint} puts additional constraints on the \vipc s.


    \item Our result is unique since we provide a characterization of all {\em randomized} single indivisible item auctions. The closest characterization result to our knowledge applies to only deterministic auctions~\citep{li2020incentive}. The example of the following \ddsic\ auction is not covered by the characterization of \citep{li2020incentive} but is covered under \Cref{thm:characterization}.
    We also provide a class of mechanisms later in this paper which subsumes many currently known mechanisms that are \ddsic.
\end{itemize}

%
\begin{figure}[t!]
    \centering
    \includegraphics[width=0.4\linewidth]{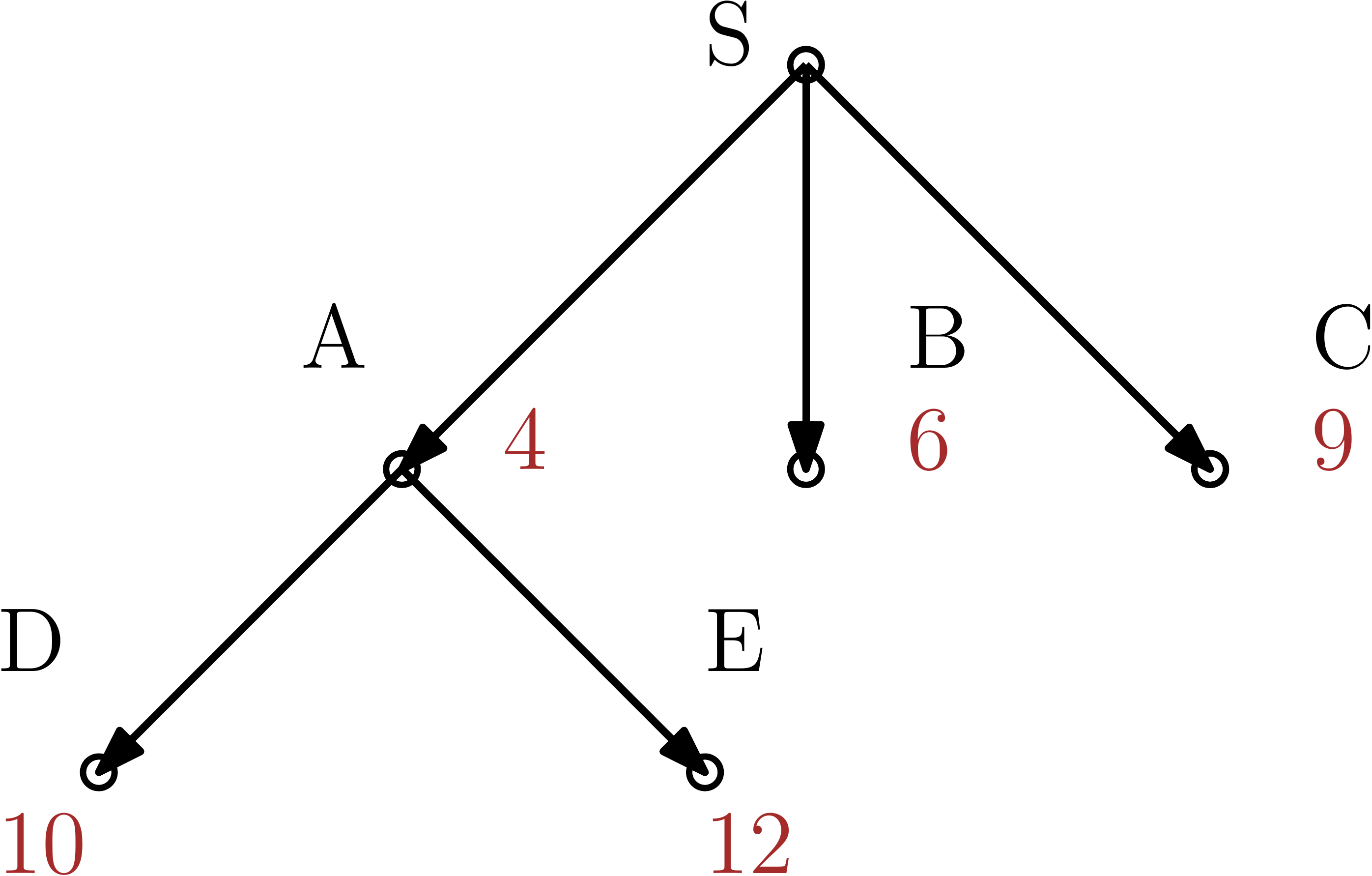}
    \caption{A randomized \ddsic\ auction.}
    \label{fig:random-ddsic}
\end{figure}

\subsection{Example to illustrate the conditions of a randomized \ddsic\ auction.}
\label{sec:example}
The distinguishing factor of the truthfulness guarantee given by \ddsic\ is in the part where an agent may not diffuse the information to its neighbors. In this example, we will focus only on that part and illustrate the meaning of the conditions of \mff{}~(\Cref{def:ffm}), that is equivalent to \ddsic{} via \Cref{cor:equivalence}.
 This example can be easily extended to a full-fledged randomized \ddsic\ auction. However, that needs the auction to be defined for {\em every} realized graph and for {\em every} type profile $((v_i,r_i),\hat{\theta}_{-i})$, which will digress a reader from the main intuition of \mff{}. Instead, we have explained how these conditions are met when the agents report their $(v_i,r_i)$s as shown in \Cref{fig:random-ddsic}.
 For simplicity of exposition, we consider the auction where the true underlying network and the reported valuations are given by \Cref{fig:random-ddsic}, and $r_i$ can take values only in $\{0,1\}$, i.e., either forward to all its neighbors or not forwarding at all.
 We discuss the satisfaction of the \mff{} conditions and consider the variation of $v_i$ and $r_i$ of each agent $i$ keeping the $\hat{\theta}_{-i}$ fixed at the values given in this figure.
 In this example, agents $D$ and $E$ get the information of the auction only if $A$ forwards it at the first level of the tree. A \ddsic\ \da\ $(g,p)$ needs to decide the allocations $g_i(f^G((v_i, r_i), \hat{\theta}_{-i}))$, and the \vipc\ components $p_i(f^G((0,r_i),\hat{\theta}_{-i}))$ for $r_i=0,1$ and for all $i \in N$.
 For all agents $i \neq A$, $r_i$'s do not matter since they do not have any children in this tree. Therefore, $g_i$'s and $\vipc_i$'s of agents $i \neq A$, remain unchanged in this example auction when they set $r_i=0$ or $r_i=1$ given other agents' reported types are fixed. Hence, condition~\ref{eq:constraint} of \mff{} is trivially satisfied for all agents except $A$. We discuss agent $A$'s satisfiability of condition~\ref{eq:constraint} separately later. In a nutshell, this example mechanism adapts the {\em residual claimant} (RC) mechanism by \citet{GL79} to this setting at the first level of the tree. If agent $A$ forwards, then it divides $A$'s probability of allocation with its children and adjusts the payments according to \Cref{def:ffm}. If $A$ does not forward, then it is just RC. Based on the forwarding decision of $A$, this example can be divided into two cases:

\begin{figure}[h!]
    \centering
    \includegraphics[width=0.5\linewidth]{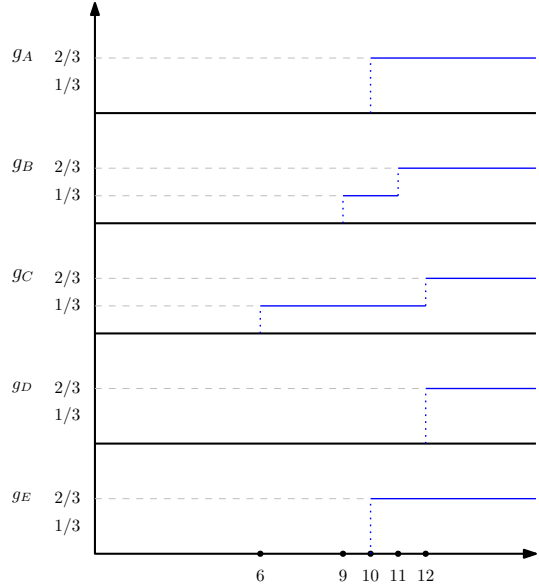}
    \caption{Allocation functions of the nodes in \Cref{fig:random-ddsic}.}
    \label{fig:allocation-example}
\end{figure}
%
 {\em Case 1: $r_A = 0$}: When agent $A$ does not forward the information, the auction stays limited to the agents $A$, $B$, and $C$. Let the auction give the object w.p.\ $2/3$ to the highest bidder and w.p.\ $1/3$ to the second highest bidder. The payment of the highest bidder is $\frac{1}{3} \times \textup{the second highest bid}$. This payment is equally distributed among the non-winning agents, which, in this case, is the third highest bidder. This is the modified version of the {\em residual claimant} mechanism~\citep{GL79}, which is DSIC (equivalent to \ddsic\ for a single-level tree).
 Hence, under this case, the allocation probability of each agent is clearly monotone non-decreasing since it increases from zero to $1/3$ when it becomes the second highest bidder and from $1/3$ to $2/3$ when it becomes the highest bidder. The \vipc\ for each agent $i$ is given by $-\frac{1}{3} \times \textup{the second highest bid in the population except agent } i$. Therefore, $\vipc_A(r_A = 0) = -6/3 = -2, \vipc_B(r_A = 0) = -4/3, \vipc_C(r_A = 0) = -4/3$. The payments follow from condition~\ref{eq:int-formula}: $p_A(r_A=0)= -2 + 0 + 0, p_B(r_A=0)= -4/3 + 6\times \frac{1}{3} -\frac{1}{3}(6-4) = 0, p_C(r_A=0)= -4/3 + 9\times \frac{2}{3} -\frac{1}{3}(6-4) - \frac{2}{3}(9-6)=2$.
 The allocation probabilities are zero for every valuation of agents $D$ and $E$, and their \vipc s are zeros. Consequently, their payments are also zero in this case.

 {\em Case 2: $r_A = 1$}:
 Let the $g_i$'s be given by \Cref{fig:allocation-example}, when agents except $i$ report their valuations as shown in \Cref{fig:random-ddsic}. Clearly, these are monotone non-decreasing. Let $\vipc_A(r_A=1) = -11/3, \vipc_B(r_A=1) = -3, \vipc_C(r_A=1) = -2, \vipc_D(r_A=1) = 0, \vipc_E(r_A=1) = 0$. The payments are given by condition~\ref{eq:int-formula} as follows: $p_A(r_A=1)= -11/3 + 0 + 0, p_B(r_A=1)= -3 + 0 + 0, p_C(r_A=1)= -2 + 9 \times \frac{1}{3} - \frac{1}{3}(9-6) = 0, p_D(r_A=1)= 0 + 0 + 0, p_E(r_A=1)= 0 + 12 \times \frac{2}{3} - \frac{2}{3}(12-10) = 20/3$.

The satisfiability of condition~\ref{eq:constraint} for agent $A$ warrants a separate discussion since it is the only agent which has a different \vipc\ when $r_A=0$ and $r_A=1$, keeping the other agents' reported types fixed. For all other agents, the satisfiability of condition~\ref{eq:constraint} is trivial since both sides of the inequality reduces to zero. However, we note that the LHS of condition~\ref{eq:constraint} for $A$ is $\vipc_A(r_A=0)-\vipc_A(r_A=1) = -2 +11/3 = 5/3$ which is larger than the RHS for every value of $v_A$. In particular, when $v_A \geqslant 10$, the RHS becomes $5/3$ and stays constant at that value for larger values of $v_A$. Hence, condition~\ref{eq:constraint} is satisfied for agent $A$ too.

This is a randomized \da\ that satisfies the \mff{} conditions (\Cref{def:ffm}) for $\hat{\theta}_{-i}$ given by \Cref{fig:random-ddsic}.
For the $r_A = 1$ case, we could have chosen any monotone allocation rule for the agents and decided the payments according to condition~\ref{eq:int-formula}, and set arbitrary \vipc{} terms for agents except $A$. But for $A$, we need to ensure that the differences in the \vipc{}s between $r_A = 0$ and $r_A = 1$ satisfies condition~\ref{eq:constraint}. This is the recipe for extending this example for every $\hat{\theta}_{-i}$.

\shortversion{
\begin{psketch}[of \Cref{thm:characterization}]
 The forward direction of the proof proceeds in two parts. First, we use the points~\ref{point1} and \ref{point2} of \ddsic, along with a few known results from convex analysis to show that the allocation function is a subgradient of the utilities of the agents and that the utilities are convex. Therefore, the allocation, being a subgradient of a convex function, is monotone non-decreasing (condition~\ref{cond1} of \Cref{thm:characterization}). Second, using the integral formula for convex functions expressed in terms of its subgradient, we extract the payment formula (\Cref{eq:int-formula}). Finally, using this payment formula in the third condition of \ddsic, we obtain \Cref{eq:constraint}, the last condition of \Cref{thm:characterization}.

 The reverse direction of the proof is straightforward, since the conditions of \Cref{thm:characterization} provide the functional forms of the payments. The proof performs a careful algebra with them to show the three points of \ddsic\ are satisfied.
\end{psketch}
}

\begin{proof}[of \Cref{thm:characterization}]
We begin with an IC mechanism $(g,p)$, and show that the conditions of the theorem hold for this mechanism.

\paragraph{IC implies condition~\ref{cond1}: the monotonicity of allocation functions.}
By definition, IC holds for all $v_i'$, and $r_i' \subseteq r_i$. Therefore, in particular, it must hold for $r_i' = r_i$. We get the following inequality:
\begin{equation} \label{eqn: ineq1}
v_ig_i(f^G((v_i,r_i),\hat\theta_{-i}) - p_i(f^G((v_i,r_i),\hat\theta_{-i}) \geq
v_ig_i(f^G((v_i',r_i),\hat\theta_{-i}) - p_i(f^G((v_i',r_i),\hat\theta_{-i})
\end{equation}
\\
Adding and subtracting $v_i'g_i(f^G((v_i',r_i),\hat\theta_{-i})$ on the RHS of \Cref{eqn: ineq1}, we get:
\begin{align*}
\lefteqn{v_ig_i(f^G((v_i,r_i),\hat\theta_{-i})) - p_i(f^G((v_i,r_i),\hat\theta_{-i}))} \\
& \geq  v_i'g_i(f^G((v_i',r_i),\hat\theta_{-i})) + (v_i-v_i')g_i(f^G((v_i',r_i),\hat\theta_{-i})) - p_i(f^G((v_i',r_i),\hat\theta_{-i})) \\
\implies & u_i((v_i,r_i), \hat\theta_{-i}) \geq u_i((v_i',r_i), \hat\theta_{-i}) + (v_i-v_i')g_i(f^G((v_i',r_i),\hat\theta_{-i}))
\end{align*}

From convex analysis \citep{rockafellar2015convex}, we know that the above inequality implies that $g_i(f^{G}((v^{\prime}_i,r_i),\hat{\theta}_{-i})$ is a sub-gradient of $u_i$ at $v^{\prime}_i$, if $u_i$ can be shown to be convex in $v_i$ for every $r_i, \hat{\theta}_{-i}$, for every $i \in N$. In the following, we show that it is indeed true.

For brevity, we use the shorthand $h(v_i) := u_i((v_i,r_i),\hat{\theta}_{-i})$
and $\phi(v_i') := g_i(f^{G}((v^{\prime}_i,r_i),\hat{\theta}_{-i}))$. Because $v_i$ and $v_i'$ were arbitrary in the above inequality, we can choose arbitrary $x_i,z_i \in {\cal V}_i$ and define $y_i={\lambda}x_i+(1-\lambda)z_i$ where $\lambda \in [0,1]$. From the above inequality, we get
\begin{align}
    h(x_i) &\geqslant h(y_i) +\phi(y_i)(x_i-y_i) \label{eqn_conv_1} \\
    h(z_i) &\geqslant h(y_i) +\phi(y_i)(z_i-y_i). \label{eqn_conv_2}
\end{align}
Multiplying \Cref{eqn_conv_1} by $\lambda$ and \Cref{eqn_conv_2} by $(1-\lambda)$ and adding, we get ${\lambda}h(x_i)+(1-\lambda)h(z_i) \geqslant h(y_i)$, which proves that $h$ or the utility $u_i$ is convex, and $\phi$ or the allocation $g_i$ is its sub-gradient. Since sub-gradient of a convex function is non-decreasing, we get the claimed implication.

\paragraph{IC implies conditions~\ref{ffm:a} and \ref{ffm:b}: the payment formulae.}

Again from convex analysis, we know that for any convex function $h$ having subgradient $\phi$, the following integral relation holds:
$h(y)=h(z)+\int_z^{y}\phi(t)dt$ for any $y,z$ in the domain of $h$. Therefore, using the same definitions of $h$ and $\phi$ from the previous case, we get (when $i$ diffuses to $r_i$)
\begin{gather}
\begin{align*}
   & u_i(f^{G}((v_i,r_i),\hat{\theta}_{-i}))
    = u_i(f^{G}((0,r_i),\hat{\theta}_{-i}))+\int_0^{v_i} g_i(f^{G}((t,r_i),\hat{\theta}_{-i}))dt \\
    \Rightarrow \ & v_ig_i(f^{G}((v_i,r_i),\hat{\theta}_{-i}))-p_i(f^{G}((v_i,r_i),\hat{\theta}_{-i}))
=-p_i(f^{G}((0,r_i),\hat{\theta}_{-i}))+\int_0^{v_i}g_i(f^{G}((t,r_i),\hat{\theta}_{-i}))dt \\
    \Rightarrow \ & p_i(f^{G}((v_i,r_i),\hat{\theta}_{-i}))
=p_i(f^{G}((0,r_i),\hat{\theta}_{-i}))+v_ig_i(f^{G}((v_i,r_i),\hat{\theta}_{-i}))-\int_0^{v_i} g_i(f^{G}((t,r_i),\hat{\theta}_{-i}))dt
\end{align*}
\end{gather}
This is precisely \Cref{eq:int-formula}, which is condition~\ref{ffm:b} of \mff{} (\Cref{def:ffm}).
%
%
%
To prove condition~\ref{ffm:a} of \Cref{def:ffm}, we first put $v_i' = v_i$ in the definition of IC to get point~\ref{point2} of \Cref{def:DDSIC} and we substitute the payment expressions derived above to get
\begin{gather}
\begin{align*}
    & \int_0^{v_i} g_i(f^G((y,r_i),\hat{\theta}_{-i})) dy - p_i(f((0,r_i),\hat{\theta}_{-i})) \geqslant \int_0^{v_i}  g_i(f^G((y,r_i'),\hat{\theta}_{-i})) dy - p_i(f((0,r_i'),\hat{\theta}_{-i})) \\
    \Rightarrow \quad & p_i(f((0,r_i'),\hat{\theta}_{-i})) - p_i(f((0,r_i),\hat{\theta}_{-i})) \geqslant \int_0^{v_i}  \left ( g_i(f^G((y,r_i'),\hat{\theta}_{-i})) - g_i(f^G((y,r_i),\hat{\theta}_{-i})) \right ) dy.
\end{align*}
\end{gather}
The first inequality follows directly where the terms $v_i g_i(f^G((v_i,r_i),\hat{\theta}_{-i}))$ cancels out on the LHS and $v_i g_i(f^G((v_i,r_i'),\hat{\theta}_{-i}))$ cancels out on the RHS. The second inequality follows by rearranging the first. Hence, this proves \Cref{eq:constraint}, condition~\ref{ffm:a} of \Cref{def:ffm}. Hence, conditions~\ref{cond1}, \ref{ffm:a}, and \ref{ffm:b} hold for a mechanism that is Incentive Compatible.
\end{proof}

We now show that \mff{} implies \ddsic{} .
\begin{theorem}[\mff{} $\Rightarrow$ \ddsic{}]
\label{thm:reverse_characterization}
If a \da{} $(g,p)$ is \mff{}, then it is \ddsic{}.
\end{theorem}

\begin{proof}
\paragraph{Conditions~\ref{cond1} and \ref{ffm:b} of \Cref{def:ffm} $\Rightarrow$ point~\ref{point1} of \ddsic.}
We are given that $g_i(f^{G}((v_i,r_i),\hat{\theta}_{-i}))$ is monotone non-decreasing in $v_i$, for a diffusion type $r_i$ (i.e. when agent $i$ diffuses to neighbour set $r_i$) and payment is given by \Cref{eq:int-formula}, for all $i \in N$.
Assuming $v_i$ to be the true valuation of agent $i$, the utility of agent $i$ when she is truthful is given by
\begin{gather}
\label{util_true}
\begin{align}
    \lefteqn{ v_ig_i(f^{G}((v_i,r_i),\hat{\theta}_{-i}))-p_i(f^{G}((v_i,r_i),\hat{\theta}_{-i}))} \\
       &= v_ig_i(f^{G}((v_i,r_i),\hat{\theta}_{-i}))-p_i(f^{G}((0,r_i),\hat{\theta}_{-i}))-v_ig_i(f^{G}((v_i,r_i),\hat{\theta}_{-i}))+\int_0^{v_i}g_i(f^{G}((t,r_i),\hat{\theta}_{-i}))dt, \nonumber
\end{align}
\end{gather}
and the utility when she misreports to $v_i'$ is given by
\begin{gather} \label{util_false}
\begin{align}
    \lefteqn{ v_ig_i(f^{G}((v_i',r_i),\hat{\theta}_{-i}))-p_i(f^{G}((v_i',r_i),\hat{\theta}_{-i}))}\\
       &= v_ig_i(f^{G}((v_i',r_i),\hat{\theta}_{-i}))-p_i(f^{G}((0,r_i),\hat{\theta}_{-i}))-v_i'g_i(f^{G}((v_i,r_i),\hat{\theta}_{-i}))+\int_0^{v_i'}g_i(f^{G}((t,r_i),\hat{\theta}_{-i}))dt. \nonumber
\end{align}
\end{gather}
Subtracting \Cref{util_false} from \Cref{util_true}, we get
\begin{gather}
  \label{res_eq}
\begin{align}
& v_ig_i(f^{G}((v_i,r_i),\hat{\theta}_{-i}))-p_i(f^{G}((v_i,r_i),\hat{\theta}_{-i}))-[v_ig_i(f^{G}((v_i',r_i),\hat{\theta}_{-i}))-p_i(f^{G}((v_i',r_i),\hat{\theta}_{-i}))] \nonumber\\
       & \quad = (v_i'-v_i)g_i(f^{G}((v_i',r_i),\hat{\theta}_{-i}))+\int_{v_i'}^{v_i}g_i(f^{G}((t,r_i),\hat{\theta}_{-i}))dt.
\end{align}
\end{gather}
Since $g_i$ is monotone non-decreasing and non-negative, the RHS of \Cref{res_eq} is always non-negative. Hence, we have point~\ref{point1} of \ddsic
\paragraph{Condition~\ref{ffm:a} $\Rightarrow$ point~\ref{point2} of \ddsic.}
Here we have \Cref{eq:constraint} satisfied. Given that we also have the expression of the payment given by \Cref{eq:int-formula} satisfied, adding and subtracting $v_i g_i(f^G((v_i,r_i),\hat{\theta}_{-i}))$ on the LHS of \Cref{eq:constraint} and adding and subtracting $v_i g_i(f^G((v_i,r_i'),\hat{\theta}_{-i}))$ on the RHS and then rearranging, we get point~\ref{point2} of \ddsic.
\end{proof}

Our final result in this section is that our definition of truthfulness, \ddsic{}, implies IC.
\begin{theorem}[\ddsic{} $\Rightarrow$ IC]
\label{thm:ddsic-ic}
 If a \da{} $(g,p)$ is \ddsic{}, then it is IC.
\end{theorem}

\begin{proof}
In this proof, we will exhaustively list all the cases of manipulation under \Cref{def:IC-zhao} and show that each of the inequalities is implied by the conditions of \ddsic\ (\Cref{def:DDSIC}). Suppose, $(v_i,r_i)$ is the tuple of the {\em true} valuation and neighbor set of agent $i$. The following cases of manipulation in $v_i$ and $r_i$ are exhaustive for \Cref{def:IC-zhao}.
\begin{itemize}
    \item Case 1: $(v_i,r_i')$, i.e., valuation is truthfully reported but diffusion is strategized.
    So, for $v_i^{\prime}=v_i$ $r_i'\subseteq r_i$, the inequality of \Cref{def:IC-zhao} becomes
\begin{align*}
          v_i g_i(f^G((v_i,r_i), \hat{\theta}_{-i})) - p_i(f^G((v_i,r_i), \hat{\theta}_{-i})) &\geqslant v_i g_i(f^G((v_i,r_i'), \hat{\theta}_{-i})) - p_i(f^G((v_i,r_i'), \hat{\theta}_{-i})) \\
         & \qquad \qquad \forall v_i, \hat{\theta}_{-i}, \forall i \in N.
\end{align*}
This is implied by condition~\ref{point2} of \Cref{def:DDSIC}.

\item Case 2: $(v_i',r_i)$, i.e., valuation is manipulated but diffusion is not. Hence, with $r_i'=r_i$, \Cref{def:IC-zhao} becomes
\begin{align*}
          v_i g_i(f^G((v_i,r_i), \hat{\theta}_{-i})) - p_i(f^G((v_i,r_i), \hat{\theta}_{-i})) &\geqslant v_i g_i(f^G((v_i',r_i), \hat{\theta}_{-i})) - p_i(f^G((v_i',r_i), \hat{\theta}_{-i})) \\
         & \qquad \qquad \forall v_i, v_i', \hat{\theta}_{-i}, \forall i \in N.
\end{align*}
This is implied by condition~\ref{point1} of \Cref{def:DDSIC}.

\item Case 3: $(v_i',r_i')$, i.e., both valuation and diffusion are strategized. Then the condition of \Cref{def:IC-zhao} becomes
    \begin{align}
          v_i g_i(f^G((v_i,r_i), \hat{\theta}_{-i})) - p_i(f^G((v_i,r_i), \hat{\theta}_{-i})) &\geqslant v_i g_i(f^G((v_i',0), \hat{\theta}_{-i})) - p_i(f^G((v_i',0), \hat{\theta}_{-i})), \nonumber \\
         & \qquad \qquad \forall v_i, v_i', \hat{\theta}_{-i}, \forall i \in N.   \label{IC_v_r}
    \end{align}
Now from condition~\ref{point2} of \Cref{def:DDSIC}, we get
    \begin{align*}
         v_i g_i(f^G((v_i,r_i), \hat{\theta}_{-i})) - p_i(f^G((v_i,r_i), \hat{\theta}_{-i})) &\geqslant v_i g_i(f^G((v_i,r_i'), \hat{\theta}_{-i})) - p_i(f^G((v_i,r_i'), \hat{\theta}_{-i})) \\
         & \qquad \qquad \forall v_i, \hat{\theta}_{-i}, \forall i \in N, \nonumber
     \end{align*}
and from condition~\ref{point1} of \Cref{def:DDSIC}, we get
\begin{align*}
         v_i g_i(f^G((v_i,r_i'), \hat{\theta}_{-i})) - p_i(f^G((v_i,r_i'), \hat{\theta}_{-i})) &\geqslant v_i g_i(f^G((v_i',r_i'), \hat{\theta}_{-i})) - p_i(f^G((v_i',r_i'), \hat{\theta}_{-i})), \\
         & \qquad \qquad \forall v_i, v_i', \hat{\theta}_{-i}, \forall i \in N.
     \end{align*}
Combining these two, we have
\begin{align*}
         v_i g_i(f^G((v_i,r_i), \hat{\theta}_{-i})) - p_i(f^G((v_i,r_i), \hat{\theta}_{-i})) &\geqslant v_i g_i(f^G((v_i',r_i'), \hat{\theta}_{-i})) - p_i(f^G((v_i',r_i'), \hat{\theta}_{-i})), \\
         & \qquad \qquad \forall v_i, v_i', \hat{\theta}_{-i}, \forall i \in N,
\end{align*}
which is \Cref{IC_v_r}.
\end{itemize}
This completes the proof.
\end{proof}

 Consolidating the results of \Cref{thm:characterization,thm:reverse_characterization,thm:ddsic-ic}, we get the following corollary.

 \begin{corollary}
 \label{cor:equivalence}
 \ddsic{} $\iff$ IC $\iff$ \mff{}.
 \end{corollary}

At the end of \Cref{sec:desiderata}, we discussed why \ddsic{} is introduced despite its equivalence with IC.

Naturally, known network auctions like IDM~\citep{li2017mechanism}, TNM~\citep{li2022diffusion}, etc., are also \ddsic{}.
In the following section, we present an unusual class of \ddsic{} auctions to demonstrate how it differs from the classical single-item auction without forwarding constraints.

\section{A non-trivial \ddsic{} mechanism}
\label{sec:examples}

We begin with a novel auction on a tree to show that there exists mechanisms which are not explored yet in the network auction parlance. We call this auction {\em Level-by-Level Exponential Valuation} (\lev) auction. This auction assigns some exponents to the agents and proceeds level-by-level from the root to the leaves.

\shortversion{There are few other auctions over networks, e.g. the Information Diffusion Mechanism (\idm) \citep{li2017mechanism} and the Threshold Neighborhood Mechanism (TNM) \citep{li2022diffusion}, which can be proved to belong to the rather stricter class of \ddsic. Due to the paucity of space, we defer the proof to the supplementary material.}

\subsection*{Level-by-Level Exponential Valuation (\lev) mechanism}
\label{sec:lblev}

We first present the high-level idea of our mechanism before formalizing it in \Cref{alg:lblev}. As the name suggests, the mechanism is run at every level of the tree $\hat{T}$ rooted at $s$ (induced by the agent reports) from the root towards the leaves. The mechanism \lev\ is parametrized by a vector $t \in \mathbb{R}_{>0}^n$, where $t_i$ denotes the {\em exponent} of agent $i$. Different choices of $t$ and $\hat{T}$ for the same input instance create a class of mechanisms, and we call each of them \lev. Let $\hat{T}_i$ be the subtree rooted at node $i$ including $i$. At each level of this tree, the mechanism sets an {\em offset} for the parent node(s) of that level, and finds the maximum valuation $v^{\max}_i$ in $\hat{T}_i$, for every $i$ at that level. It deducts the offset from $v^{\max}_i$ to calculate $i$'s {\em effective valuation} $\rho_i$ and decides which $\hat{T}_i$'s ``stay in the game''. For the $i$'s that stay, the mechanism considers the largest $\rho_i^{t_i}$ and tentatively sets it as the ``winning subtree'' and calculates its ``actual payment''. The mechanism repeats at every next level treating the current root of the tentatively winning subtree as the parent with an updated offset. \Cref{alg:lblev} details out the description of this mechanism in an algorithmic manner.
\begin{algorithm}[t]
\caption{\lev}
\label{alg:lblev}
 \KwIn{reported types $\hat{\theta}_i = (\hat{v}_i, \hat{r}_i), \hat{r}_i \subseteq r_i, \ \forall i \in N$}
 \KwOut{winner of the auction (which can be $\emptyset$), payments of each agent}
 {\bf Preprocessing:} Since the underlying graph is a tree, let $\hat{T}$ be the sub-tree rooted at $s$ induced from $\hat{r}_i, i \in N$.
 Pick $t\in\mathbb{R}_{> 0}^n$ independent of the input\; \label{line-prep}
\uIf{$\hat{v}_i = 0, \forall \ i \in N$\label{line00}}
    {Item is not sold and payment is set to zero for all agents, STOP\;}
{\em Initialization:} all agents are \texttt{non-winners} and their \texttt{actual payments} are zeros, set \texttt{offset} = 0, \texttt{level} = 1, \texttt{parent} = $s$, $v_\pa = 0$\;
In this level of $\hat{T}$: \; \label{stepfirst}
\Indp
\For{each node $i \in \ch(\pa)$}{
    Set \texttt{effective valuation} $\rho_i := \max\{\hat{v}_j : j \in \hat{T}_i\} - \of$\;
}
Remove the nodes that have $\rho_i < 0$, denote the rest of the agents with $N_\textup{\tt remain}$\; \label{remove}
\uIf{$|N_\textup{\tt remain}| \geqslant 2$}{
Sort the nodes in decreasing order of $\rho_i^{t_i}$ \;
Compute $z := \rho_\ell^{t_\ell / t_{i^*}}$, where $i^*$ is the highest in this order and $\ell$ is the second highest node in the decreasing $\rho_i^{t_i}$ order\;
}
\uElse{
Set $z=0$\;
} \label{line09}
\uIf{$v_{\pa}\geqslant \of +z$\label{line10}}
    {STOP and go to Step~\ref{finalstep} \;}
Set the highest node $i^*$ in this order as the \texttt{tentative winner} and its \texttt{effective payment} to be $z$ \;\label{line13}
All nodes and their subtrees except $i^*$ are declared \texttt{non-winners} \;
The \texttt{actual payment} of $i^*$ to \texttt{parent} $=$ \texttt{effective payment} + \texttt{offset} \;
\texttt{parent} = $i^*$, $\texttt{offset} =  \text{ \texttt{actual payment} of } i^* $ \; \label{steppayment}
\texttt{level} = \texttt{level} + 1 \; \label{steplast}
\Indm
Repeat Steps~\ref{stepfirst} to \ref{steplast} with the updated \texttt{parent} and \texttt{offset} for the new level \;
STOP when no agent $i$ has $\rho_i \geqslant 0$ OR the leaf nodes are reached \; \label{line18}
Set \texttt{tentative winner} as \texttt{final winner}; final payments are the \texttt{actual payments} that are paid to the respective parents of $\hat{T}$ \label{finalstep}
\end{algorithm}

In the sub-tree $\hat{T}$, every agent receives the difference between the payment that their children in $\hat{T}$ give to that agent and the payment she makes to her parent.
The algorithm terminates either at some leaf node or at a node that has large enough offset such that none of its children `stay in the game'. We call that node the winner of \lev.


\longversion{
\noindent
{\bf Illustration of \lev\ through an example.}
Suppose the sub-tree $\hat{T}$ generated from the reported graph $\hat{G}$ is as shown in \Cref{fig:lblev} for $11$ nodes (named $A, B, \ldots, K$ in the figure) in addition to the seller $S$. The tuple next to agent $i$ denotes $(v_i, t_i)$, for all $i = A, B, \ldots, K$, where $v_i$ is the reported valuation and $t_i$ is the exponent set by the mechanism.
\begin{figure}[h!]
    \centering
    \includegraphics[width=0.7\textwidth]{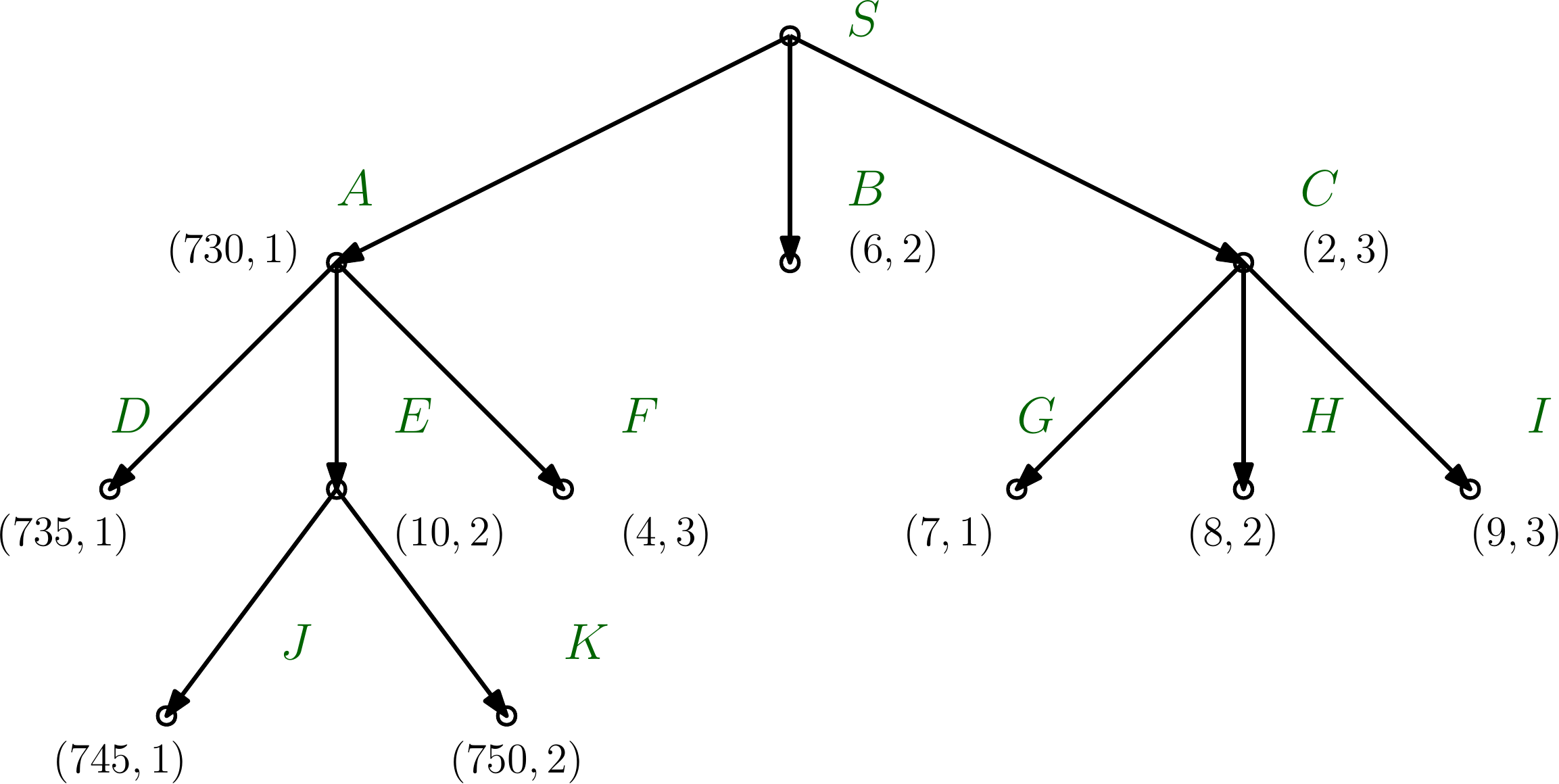}
    \caption{An example instance of \lev.}
    \label{fig:lblev}
\end{figure}
At {\tt level} = 1 of $\hat{T}$, from \Cref{alg:lblev} we get \of = 0. We find the {\tt effective valuations} to be $\rho_A=750$, $\rho_B=6$, $\rho_C=9$. Now, we observe that ${\rho_A}^{t_A}=750$ is the highest among the ${\rho_i}^{t_i}$s of that level. Hence, by line~\ref{line13}, agent $A$ is set as the {\tt tentative winner} and agent $B$ and $C$ and their subtrees are set as {\tt non-winners}.
Also, the {\tt effective payment} of agent $A$ to $S$ is $\rho_B^{t_B/t_A}=9^{3/1}=729$.
From line~\ref{steppayment}, the {\tt actual payment} of $A$ becomes the same as its {\tt effective payment} since at this level \of = 0. Also, for the next iteration, i.e., for {\tt level} = 2, $\pa=A$ and $\of=729$ are set.

At {\tt level} = 2, $\rho_D=735-729=6$, $\rho_E=750-729=21$ and $\rho_F=4-729=-725$. As $\rho_F < 0$, $F$ and its subtree is removed (line~\ref{remove}). Line~\ref{line10} stands false, since $v_A = 730 \ngeqslant 729 + 6^{1/2} = \of + \rho_\ell^{t_\ell / t_{i^*}}$. Hence, $E$ is the {\tt tentative winner}, and it pays $729+6^{1/2}=731.45$.
Again from line~\ref{steppayment}, the next level ({\tt level} = 3) details are updated, i.e., $\pa=E$ and $\of=731.45$.

At {\tt level} = 3, $\rho_J=745-731.45=13.55$, $\rho_K=750-731.45=18.55$, Agent $K$ becomes the {\tt tentative winner} as line~\ref{line10} returns false. The {\tt actual payment} of $K$ is $731.45+13.55^{1/2}=735.13$.

As agent $K$ is a leaf node, the algorithm stops via line~\ref{line18} and $K$ is declared as the {\tt final winner} (line~\ref{finalstep}). Agent $A$ gets the difference between the amounts it receives from its children and pays to its parent, i.e., $731.45-729=2.45$ (one can think of this amount as the {\em commission} to forward the information, which offsets its payoff when it manipulates and does not forward). Similarly, agent $E$ gets $3.681$ as the commission.
The revenue generated by the auction is $729$ which is the payment of agent $A$ to $S$.
}

%
Note that, \lev\ actually defines a class of mechanisms. A specific instance of an \lev\ mechanism is identified by the chosen vector $t$. However, keeping with the tradition of the mechanism design literature, we call all such mechanisms \lev. For example, two instances of Groves mechanism may have different $h_i(\theta_{-i}), i \in N,$ functions, but both of them are known by the same name.
In the following results, we provide two important properties of \lev.

\begin{theorem}
\label{thm:lblev-truthful}
 \lev\ is \ddsic.
\end{theorem}

\longversion{
\noindent
{\em Remark}: As discussed earlier in \Cref{sec:results}, in the classical single object auction characterization result by \citet{Myerson1981}, the \vipc\ term could have been chosen independently. For every such choice, the mechanism would have been DSIC in that setting. In the case of an auction on the network, we need to additionally ensure the diffusion constraint (\Cref{eq:constraint}) that restricts the choice of the \vipc s. We show that the \vipc s corresponding to the \lev\ mechanism ensure it and this is where these mechanisms on the network are distinctly different from that of Myerson's characterization.
}

\shortversion{
\begin{psketch}
 The proof proceeds by showing that \Cref{alg:lblev} satisfies all the conditions of the characterization result (\Cref{thm:characterization}) -- in particular, the diffusion constraint (\Cref{eq:constraint}) that restricts the choice of the \vipc s (unlike \citet{Myerson1981} where \vipc s could have been chosen independently). Given an instance of the reported types, we partition the agents into {\em three} exhaustive classes: (i) \win\ -- agent who gets the item (\lev\ is deterministic, hence there will be a deterministic winner), (ii) \opnw\ -- agents that lie on the path from the seller to the winner, and (iii) \nopnw\ -- agents that are not on the winning path from the seller to the winner. We show that the allocation function satisfies cond.~\ref{cond1} of \Cref{thm:characterization} for all these agents. The payments given by \Cref{alg:lblev} matches \Cref{eq:int-formula} for each of them with appropriate choices of the \vipc\ terms identified from the algorithm. The most crucial part of the proof is to show that the \Cref{eq:constraint} is also satisfied by the \vipc\ terms. The rest of the proof shows that the offset chosen by the algorithm at each node and updating the {\em effective} valuations of the nodes in its subtree or eliminating the branch of the subtree that does not have any node having a valuation at least as much as the offset ensures this critical diffusion constraint.
\end{psketch}
}

Before embarking on the proof, we want to give an overview since it is long and detailed and also define a few terminologies. The proof proceeds by showing that \Cref{alg:lblev} satisfies all the three conditions of \mff{}, which is equivalent to \ddsic{} (\Cref{cor:equivalence}). Given an instance of the reported types, we partition the agents into {\em three} classes that are exhaustive: (i) \win\ -- agent who gets the item (\lev\ is deterministic, hence there will be a deterministic winner), (ii) \opnw\ -- agents that lie on the path from the seller to the winner, and (iii) \nopnw\ -- agents that are not on the winning path from the seller to the winner. We show that the allocation function satisfies cond.~\ref{cond1} of \mff{} (\Cref{def:ffm}) for all these agents. The payments given by \Cref{alg:lblev} matches \Cref{eq:int-formula} for each of them with appropriate choices of the \vipc\ terms identified from the algorithm. The most crucial part of the proof is to show that the \Cref{eq:constraint} is also satisfied by the chosen \vipc\ terms. The \of\ chosen by the algorithm at each {\tt level} is crucial. Updating the {\em effective} valuations of the nodes in the subtree of the current \pa\ and elimination of the branch of the subtree that has all valuations smaller than the \of, ensure this critical diffusion constraint (\Cref{eq:constraint}).

\begin{proof}[of \Cref{thm:lblev-truthful}]
 The proof is divided into three parts, each of which shows that \lev\ satisfies the three conditions of \mff{} (\Cref{def:ffm}). Since \lev\ is defined in a manner where sub-auctions happen in every level of the sub-tree $\hat{T}$ obtained from $\hat{G}$, we need a few terms defined for a cleaner presentation of the proof. These are:
 \begin{itemize}
     \item $\of(i)$: $\of$ for the $\ch$ of $i$, i.e., at a level where $i$ is the auctioneer.
     \item $\ch(i)$ and $\pa(i)$ are the $\ch$ and $\pa$ of $i$ respectively in $\hat{T}$, and is defined in the usual way for standard trees. Therefore, $\of(\pa(i))$ will denote the $\of$ set at one level before agent $i$ by the $\pa$ of $i$, and will refer to the previous iteration of the \lev\ mechanism. Similarly, $\ch(\pa(i))$ denotes the siblings of $i$ including herself.
     \item $\wi(i):= \argmax_{j \in \ch(i)} \rho_j^{t_j}$, and $\ru(i):= \argmax_{j \in \ch(i) \setminus \wi(i)} \rho_j^{t_j}$ denote the $\wi$ and $\ru$ respectively of the auction at a level where $i$ is the auctioneer. Ties are broken arbitrarily in both these cases.
 \end{itemize}

In the extreme case where the reported valuation of every agent is zero, the allocation probability is zero for every agent and so are their payments (according to \cref{line00} of \Cref{alg:lblev}). Here it trivially satisfies all the three conditions of \Cref{def:ffm}. So, in the rest of the proof, we will assume that at least one agent has a positive reported valuation. Therefore, if some agent reports her valuation to be zero, she is not allocated the object.

\paragraph{Part 1: \lev\ satisfies Condition~\ref{cond1} of \mff{}:}

To show that the allocation function under \lev\ satisfies monotonicity w.r.t.\ the valuation of every agent, we need to show that for every pair $v_i,v_i'$ s.t. $v_i' > v_i$, the allocation probability at $v_i'$ is at least as much as at $v_i$, for all $r_i$ (when agent $i$ diffuses to $r_i$), and for all $i \in N$. Since the \lev\ mechanism is defined w.r.t.\ the effective valuations $\rho_i$'s, and not $v_i$'s, showing this is non-trivial. Based on the fact that a typical agent $i$ can belong to one of the three classes after the outcome of the mechanism is chosen, we have the following cases when agent $i$ reports a valuation of $v_i$.

\smallskip \noindent
{\em Case 1:} agent $i$ is a \nopnw: in this case, $g_i(f^{G}(v_i,r_i),\hat{\theta}_{-i}) = 0$. From the description of \lev, it is clear that for $v_i' > v_i$, either agent $i$ can remain a \nopnw, or it can become a \win. It cannot become an \opnw\ because it would imply that there was another agent in $i$'s subtree that had a maximum valuation in this network at agent $i$'s original valuation $v_i$ and then agent $i$ could not be a \nopnw. In both the cases where agent $i$ is \nopnw\ or \win, $g_i(f^{G}(v_i',r_i),\hat{\theta}_{-i}) \geqslant 0$, hence it is monotone non-decreasing.

\smallskip \noindent
{\em Case 2:} agent $i$ is an \opnw: the allocation for agent $i$ is  $g_i(f^{G}(v_i,r_i),\hat{\theta}_{-i}) = 0$ in this case as well. This agent is \opnw\ with bid $v_i$ implies that there is an agent in $\hat{T}_i$ that has reported the winning bid. Now, if agent $i$ bids $v_i'$ which is higher than $v_i$, it can either continue to be an \opnw\ or may become the new \win\ at a sufficiently high bid. In both these cases, the allocation probability is monotone non-decreasing.

\smallskip \noindent
{\em Case 3:} agent $i$ is the \win: here $g_i(f^{G}(v_i,r_i),\hat{\theta}_{-i}) = 1$. We need only to show that for all $v_i' > v_i$, agent $i$ continues to be the winner. This is fairly easy to see from \Cref{alg:lblev}. An agent can be the winner either when it is the parent node in \cref{line10} or \cref{line18}.

In \cref{line10}, since agent $i$ is the parent and it satisfies the {\tt if} condition of that line, an increase in its valuation will continue to hold that condition true and $i$ will continue to be the \win.

In \cref{line18}, agent $i$ is the auctioneer whose \of\ is higher than the valuations of all agents in its subtree. The \of\ is not a function of agent $i$'s valuation -- it is determined by $\rho_\ell^{t_\ell/t_i}$, where $\ell$ is the second highest node in the decreasing $\rho_k^{t_k}$ order (\cref{line09}). Hence, an increase in $v_i$ will continue keeping $i$ to be the winner.

These three cases together prove this part of the proof.

\paragraph{Part 2: \lev\ satisfies Condition~\ref{ffm:b} of \mff{}:}

To show this part of the proof, we need to show that the payments given by \Cref{alg:lblev} for each agent matches \Cref{eq:int-formula}. Note that, after a monotone non-decreasing allocation rule has been picked (as seen in the previous case of this proof) by the algorithm, the only variable quantity in the payment formula is the \vipc\ term. The other two terms, i.e., the second and third term on the RHS of \Cref{eq:int-formula} are already fixed given the allocation. So, in order to complete the proof, we need to find an appropriate \vipc\ such that the payment given by \lev\ exactly matches the sum of those two terms and the \vipc.\footnote{Since we discuss the values of the \vipc s at a specific instance, i.e., after the variables $r_i', \hat{\theta}_{-i}$ have realized, the arguments of those \vipc\ terms are clear from the context. Therefore, we will omit the arguments of the \vipc\ terms for brevity in the rest of the paper.} We will denote the payment for agent $i$ given by the mechanism as $p^{\lev}_i((v_i,r_i'),\hat{\theta}_{-i})$ when the reported type of agent $i$ is $(v_i,r_i')$ and that for the other agents are $\hat{\theta}_{-i}$.


\smallskip \noindent
{\em Case 1:} agent $i$ is a \nopnw: note that, in this case, $g_i(f^{G}(v_i,r_i'),\hat{\theta}_{-i}) = 0$. Hence, the last two terms in the RHS of \Cref{eq:int-formula} vanishes. We need to set that the $\vipc_i$ term which is exactly equal to $p^{\lev}_i((v_i,r_i'),\hat{\theta}_{-i})$, the actual payment of agent $i$ under \lev, and show that it is indeed independent of $v_i$ for it to be qualified as a \vipc. From the algorithm, we see that $p^{\lev}_i((v_i,r_i'),\hat{\theta}_{-i}) = 0$, since $i$ is a \nopnw\ agent. Hence, $\vipc_i=0$, and it matches conditions~\ref{ffm:a} and \ref{ffm:b} of \Cref{def:ffm}.

\smallskip \noindent
{\em Case 2:} agent $i$ is an \opnw: for this case as well, the situation is similar to the previous case: $g_i(f^{G}(v_i,r_i'),\hat{\theta}_{-i}) = 0$, and hence, the last two terms in the RHS of \Cref{eq:int-formula} vanishes. We need to calculate $p^{\lev}_i((v_i,r_i'),\hat{\theta}_{-i})$, the actual payment of agent $i$ under \lev, and show that it is indeed independent of $v_i$ for it to be qualified as $\vipc_i$.

From the \lev\ algorithm, we see that the net payment of a \opnw\ agent $i$ has the following simple structure:
\begin{equation}
\label{eq:nopnw-commission}
 p^{\lev}_i((v_i,r_i'),\hat{\theta}_{-i}) = \pi(\hat{T}_i)-\sum_{j \in \ch(i)} \pi(\hat{T}_j) =: \pi(\hat{T}_i) - R_i, \ \forall r_i' \subseteq r_i
\end{equation}
Where $\pi(\hat{T}_k)$ is payment made by the subtree rooted at $k$ to its parent (called the \texttt{actual payment} in \Cref{alg:lblev}). Therefore, $\sum_{j \in \ch(i)} \pi(\hat{T}_j)$ is the net payment received by agent $i$ from the subtrees rooted at its children nodes. Therefore, the payment of agent $i$ is just the difference between the payment it makes to $\pa(i)$ minus the sum of the payments it receives from $\ch(i)$ (according to \cref{finalstep}). We use the shorthand $R_i$ to denote $\sum_{j \in \ch(i)} \pi(\hat{T}_j)$.

Now, we need to show that the RHS of \Cref{eq:nopnw-commission} is independent of $v_i$ and then we are done claiming it to be $\vipc_i$. From the algorithm, we find that the payment received by $i$ has a rather simpler form:
\begin{equation}
    \label{eq:received-payment}
    R_i = \of(i) + \rho_\ell^{t_\ell / t_k}, \textup{ where } k = \wi(i), \ell = \ru(i).
\end{equation}
The above equation means that the payment received by $i$ is the {\tt actual payment} the winner of this level, $k$, makes to its parent $i$, and it is the sum of two terms: (a)~the offset set by $i$ while auctioning at that level and (b)~the $\rho$ of the second highest bidder in the decreasing order of $\rho_i^{t_i}$ raised to an appropriate exponent.

From \cref{steppayment}, we find that $\of(i) = \pi(\hat{T}_i)$, i.e., the payment the winning agent $i$ of a level makes to its parent is set as the \of\ in the next level. Note that, there must be at least one next level since it is an \opnw. Therefore, the RHS of \Cref{eq:nopnw-commission} becomes $-\rho_\ell^{t_\ell / t_k}$. We claim that this is independent of $v_i$. The exponents are constants and independent of $v_i$. The term $\rho_\ell = v_\ell - \of(i)$ is also independent of $v_i$. This is because the first term $v_\ell$ is independent of $v_i$. The second term $\of(i)$ is the payment $\pi(\hat{T}_i)$, that $i$ made to its parent which is a function of the $\of(\pa(i))$ and the valuations of the agents other than agent $i$ in the level where $\pa(i)$ was the auctioneer and $i$ was a participant of that auction.
\[\pi(\hat{T}_i) = \of(\pa(i)) + \rho_j^{t_j / t_i}, \textup{ where } j = \ru(\pa(i)).\]
Since, both these terms are independent of $v_i$, the RHS of \Cref{eq:nopnw-commission} is independent of $v_i$ and hence it is a valid $\vipc_i$.

\smallskip \noindent
{\em Case 3:} agent $i$ is the \win: unlike the previous two cases, here $g_i(f^{G}(v_i,r_i'),\hat{\theta}_{-i}) = 1$. Therefore, the payment given by \Cref{eq:int-formula} has the last two terms on the RHS that are non-zero. The first term of the RHS, i.e., the \vipc, is the payment of the agent when it reports $0$ instead of $v_i$. From the algorithm, we observe that an agent $i$ can become a winner in two possible ways: (i)~if $i$'s valuation is larger than the maximum payment it can extract from $\ch(i)$ in $\hat{T}_i$ (\cref{line10}), or (ii)~if $i$'s offset is so high that none of its children has a positive {\tt effective valuation} $\rho$ or $i$ is a leaf node (\cref{line18}). In the first case, if agent $i$ reports a valuation of $0$, then it becomes a \opnw, and in the second, it becomes a \nopnw. In the following, we consider these two cases separately and show that $p^{\lev}_i((v_i,r_i'),\hat{\theta}_{-i})$ indeed matches the expression of \Cref{eq:int-formula}.
\begin{itemize}
    \item {\em Case 3(i):} bidding $0$ makes agent $i$ a \opnw: when agent $i$ is an \opnw, its payment is $\vipc_i = -\rho_\ell^{t_\ell / t_k}, \textup{ where } k = \wi(i), \ell = \ru(i)$ (from Case 2 above). Given the allocation function is already fixed via the \lev\ mechanism, the last two terms in \Cref{eq:int-formula} can be written w.r.t.\ that allocation as
    \begin{align*}
        \lefteqn{g_i(f^{G}((v_i,r_i'),\hat{\theta}_{-i}))v_i-\int_{0}^{v_i}g_i(f^{G}((v,r_i'),\hat{\theta}_{-i}))dv} \\
        &=
        g_i(f^{G}((v_i,r_i'),\hat{\theta}_{-i}))v_i-\int_{0}^{l_i}g_i(f^{G}((v,r_i'),\hat{\theta}_{-i}))dv-\int_{l_i}^{v_i}g_i(f^{G}((v,r_i'),\hat{\theta}_{-i}))dv \\
        &= 1\cdot v_i-\int_{0}^{l_i} 0\cdot dv-\int_{l_i}^{v_i}1 \cdot dv = l_i,
    \end{align*}
    where $l_i$ is the threshold after which agent $i$ starts becoming the winner. The \win\ of the \lev\ mechanism is deterministic and this agent starts becoming the \win\ when its valuation crosses a threshold point that we define to be $l_i$, which is guaranteed to exist. Now, we see that agent $i$, which was an \opnw, starts becoming a \win\ only when it is the \pa\ in \cref{line10} of the algorithm. Hence the threshold $l_i$ will be as follows.
    \[l_i = \of(i) + \rho_\ell^{t_\ell / t_k}, \textup{ where } k = \wi(i), \ell = \ru(i).\]
    Therefore, the entire payment given by \Cref{eq:int-formula} is $\vipc_i + l_i = \of(i)$.
    From the algorithm, we notice that the \win\ pays its own \of\ to its parent. Therefore,
    \begin{align*}
     p^\lev_i((v_i,r_i'),\hat{\theta}_{-i})
           = \of(i)
    \end{align*}
    Hence, we have the equality of the \lev\ payment with that of \Cref{eq:int-formula}.
    \item {\em Case 3(ii):} bidding $0$ makes agent $i$ a \nopnw: when agent $i$ is a \nopnw, its payment is $\vipc_i=0$ (from Case 1 above). Given the allocation function is already fixed via the \lev\ mechanism, the last two terms in \Cref{eq:int-formula} can be written w.r.t.\ that allocation as
    \begin{align*}
        \lefteqn{g_i(f^{G}((v_i,r_i'),\hat{\theta}_{-i}))v_i-\int_{0}^{v_i}g_i(f^{G}((v,r_i'),\hat{\theta}_{-i}))dv} \\
        &=
        g_i(f^{G}((v_i,r_i'),\hat{\theta}_{-i}))v_i-\int_{0}^{k_i}g_i(f^{G}((v,r_i'),\hat{\theta}_{-i}))dv-\int_{k_i}^{v_i}g_i(f^{G}((v,r_i'),\hat{\theta}_{-i}))dv \\
        &= 1\cdot v_i-\int_{0}^{k_i} 0\cdot dv-\int_{k_i}^{v_i}1 \cdot dv = k_i,
    \end{align*}
    where $k_i$ is the threshold after which agent $i$ starts becoming the winner. The \win\ of the \lev\ mechanism is deterministic and this agent starts becoming the \win\ when its value crosses a threshold point that we define to be $k_i$, which is guaranteed to exist. Since, in this case, $i$ was a \nopnw\ till its value reached $k_i$, this critical valuation is given by
    \[k_i = \of(\pa(i)) + \rho_j^{t_j/t_i}, \textup{ where } j=\ru(\pa(i)).\]
    If $v_i$ crosses the value on the RHS above, then $\rho_i$ becomes the maximum among the children of $\pa(i)$, and hence it becomes the winner.

    On the other hand, under this case, $i$ becomes the winner because \cref{line18} of the \lev\ mechanism becomes effective. Hence, we have
    \begin{align*}
     p^\lev_i((v_i,r_i'),\hat{\theta}_{-i})
           = \of(\pa(i))+\rho_j^{t_j/t_i}, \textup{ where } j=\ru(\pa(i))
    \end{align*}
    Since $\vipc_i=0$ in this case, we have the equality of the \lev\ payment with that of \Cref{eq:int-formula}.
\end{itemize}

\paragraph{Part 3: \lev\ satisfies Condition~\ref{ffm:a} of \mff{}:}

In this condition, we need to compare $\vipc_i$ between the cases when $i$ forwards to its complete neighbor set ($r_i$) versus a subset of its neighbors ($r_i'\subseteq r_i$). By reporting a diffusion type, agent $i$ can be in one of the three classes: \nopnw, \opnw, or \win. We handle these cases one by one.

\noindent
{\em Case 1: agent $i$ is a \nopnw\ when it diffuses to complete neighbor set $r_i$:}, then either all nodes from $i$ to the root $s$ on $\hat{T}$ were never tentative winners, or some node in that path is the winner. In both cases, if agent $i$ does not forward the information to $r_i$ and instead diffuses to $r_i'$, s.t. $r_i'\subseteq r_i$, the winner does not change. Hence, $\vipc_i = 0$ and $g_i = 0$ for any diffusion type. Therefore, \Cref{eq:constraint} is trivially satisfied.

\noindent
{\em Case 2: agent $i$ is the \win\ node when it diffuses to complete neighbor set $r_i$:} then \lev\ already treats $i$ as if it does not forward and calculates $\vipc_i$. So, $\vipc_i$ of agent $i$ at diffusion type $r_i$ is the same as when $i$ forwards to $r_i'$, some subset of $r_i$. Also, $i$ will continue to be the \win\ for any diffusion type. Hence $g_i = 1$ for any diffusion type reported by agent $i$. Therefore, similar to {\em Case 1}, \Cref{eq:constraint} is trivially satisfied.

\noindent
{\em Case 3: agent $i$ is an \opnw\ node when it diffuses to entire neighbor set $r_i$:} this is a non-trivial case, since by strategic forwarding, agent $i$ may change the winner. By partial or not forwarding, an \opnw\ agent $i$ can either become a \nopnw\ or a \win.
\begin{itemize}
    \item When $i$ becomes a \nopnw\ by diffusing to $r_i' \subset r_i$, its utility becomes $0$ (see {\em Case 1} of Part 2 above). However, an \opnw\ draws utility $R_i - \pi(\hat{T}_i)$ where $R_i = \of(i) + \rho_\ell^{t_\ell / t_k}=\pi(\hat{T}_i) + \rho_\ell^{t_\ell / t_k}$ (see {\em Case 2} of Part 2 above).
    As $\rho_\ell \geqslant 0$, hence utility being an \opnw\ is non-negative and makes diffusion to $r_i$ a weakly better option for $i$.

    \item In the other case, when $i$ becomes a \win\ by strategic forwarding, its utility becomes $v_i - \pi(\hat{T}_i)$.
    Note that $i$ is an \opnw\ because it failed the {\tt if} condition in \cref{line10}, hence,
\begin{equation}
    \label{eq:value-less}
    v_i < \of(i) + \rho_\ell^{t_\ell / t_k}, \textup{ where } k = \wi(i), \ell = \ru(i).
\end{equation}
Now, as an \opnw, $i$'s utility is $R_i - \pi(\hat{T}_i)$ (since the allocation probability of $i$ is zero by \Cref{eq:nopnw-commission}). However, the actual payment by $\ch(i)$ to $i$ is
\begin{equation}
    \label{eq:payment-more}
    R_i = \of(i) + \rho_\ell^{t_\ell / t_k}, \textup{ where } k = \wi(i), \ell = \ru(i).
\end{equation}
Therefore, from \Cref{eq:value-less,eq:payment-more} we get, $v_i < R_i$. Agent $i$ pays $\pi(\hat{T}_i)$ to its parent regardless of whether it forwards to $r_i$ or not. Therefore, if agent $i$ forwards to $r_i$, it gets an utility of $R_i - \pi(\hat{T}_i) = \rho_\ell^{t_\ell / t_k} \geqslant 0$ which is larger than the utility $v_i - \pi(\hat{T}_i)$ of $i$ when it diffuses to some $r_i'\subset r_i$ and becomes the \win. Therefore, in this case as well, forwarding to $r_i$ is better than any partial forwarding to $r_i' \subset r_i$ for agent $i$.
\end{itemize}
Since agent $i$ and $r_i'\subseteq r_i$ is arbitrary, we conclude that forwarding to the entire neighbor set $r_i$ is a weakly dominant strategy than forwarding to a subset $r_i'$ for every $i$ in each of the three cases above. This gives,
\begin{equation*}
v_ig_i(f^G((v_i,r_i),\hat{\theta}_{-i})) - p_i(f^G((v_i,r_i),\hat{\theta}_{-i})) \geqslant v_ig_i(f^G((v_i,r_i'),\hat{\theta}_{-i})) - p_i(f^G((v_i,r_i'),\hat{\theta}_{-i}))
\end{equation*}
Since, we have already shown in the previous part of this proof that the payment expression is given by \Cref{eq:int-formula}, expanding $p_i(f^G((v_i,r_i),\hat{\theta}_{-i}))$ and $p_i(f^G((v_i,r_i'),\hat{\theta}_{-i}))$ according to that expression in the above inequality, we get
\begin{equation*}
p_i(f^G((0,r_i'),\hat{\theta}_{-i})) - p_i(f^G((0,r_i),\hat{\theta}_{-i})) \geqslant \int_0^{v_i}  \left ( g_i(f^G((y,r_i'),\hat{\theta}_{-i})) - g_i(f^G((y,r_i),\hat{\theta}_{-i})) \right ) dy.
\end{equation*}
Hence, \Cref{eq:constraint} holds for \lev.

Hence, combining the Parts 1 to 3, we conclude that \lev\ is \ddsic.
\end{proof}
\begin{theorem}
\label{thm:lblev-ir}
 \lev\ is IR.
\end{theorem}

\shortversion{
\begin{psketch}
 This conclusion is straightforward for \nopnw\ agents since both their allocation and payment are zero. For the \opnw\ agents, the allocation probability is zero and we show that their received payment from the subtree below is at least as much as their payment to the parent node. The critical part of the proof is for the \win\ agents. We need to consider the cases under which a \win\ wins the auction (decided based on its effective valuation), and show that the difference between its valuation and payment under \lev\ is non-negative.
\end{psketch}
}

\longversion{
\begin{proof}
 We need to show that for every agent $i \in N$, the net utility $v_i - p^{\lev}_i((v_i,r_i),\hat{\theta}_{-i}) \geqslant 0$. We know that for a reported type profile $((v_i,r_i),\hat{\theta}_{-i})$, agent $i$ can be one of the following.
 \begin{itemize}
     \item Agent $i$ is \nopnw: in this case, $i$ has an allocation probability of zero, since it never becomes a \texttt{tentative winner}. Also, by \Cref{alg:lblev}, its payment remains zero throughout. Therefore, such an agent satisfies the IR condition (\Cref{def:IR}) trivially.
     \item Agent $i$ is \opnw: in this case, $i$'s allocation probability is still zero, and it makes a payment of $\pi(\hat{T}_i) - R_i$ as given by \Cref{eq:nopnw-commission}. In the discussion following that equation, we see that $R_i = \of(i) + \rho_\ell^{t_\ell / t_k}, \textup{ where } k = \wi(i), \ell = \ru(i),$ (\Cref{eq:received-payment}) and that $\of(i) = \pi(\hat{T}_i)$. Therefore, the utility of agent $i$ in this case is $0 - \pi(\hat{T}_i) + R_i = \rho_\ell^{t_\ell / t_k} \geqslant 0$.
     \item Agent $i$ is \win: from the algorithm, we observe that an agent $i$ can become a winner in two possible ways: (i)~if $i$'s valuation is larger than the maximum payment it can extract from $\ch(i)$ in $\hat{T}_i$ (\cref{line10}), or (ii)~if $i$'s offset is so high that none of its children has a positive {\tt effective valuation} $\rho$ or $i$ is a leaf node (\cref{line18}). However, in both the cases, the payment of the agent is given by $\of(\pa(i)) + \rho_j^{t_j/t_i}$, where $j = \ru(\pa(i))$.

     Also, agent $i$ is the winner implies that its $\rho_i$ is such that $\rho_i^{t_i}$ is the largest among all its siblings, i.e., $\ch(\pa(i))$. Therefore, we can write
     \begin{align*}
         & \rho_i^{t_i} \geqslant \rho_j^{t_j}, \textup{ where } j = \ru(\pa(i)) \\
         \Rightarrow \qquad & \rho_i \geqslant \rho_j^{t_j/t_i}, \textup{ since } \rho_i, t_i > 0 \\
         \Rightarrow \qquad & v_i \geqslant \of(\pa(i)) + \rho_j^{t_j/t_i} = p^{\lev}_i((v_i,r_i),\hat{\theta}_{-i})\\
         \Rightarrow \qquad & v_i - p^{\lev}_i((v_i,r_i),\hat{\theta}_{-i}) \geqslant 0.
     \end{align*}
     The second implication follows from the fact that \lev\ sets $\rho_i$ to be the difference between the reported valuation of $i$ and the offset set by its parent node. Therefore, the RHS becomes the payment of agent $i$.
 \end{itemize}
 Considering the three cases above, we conclude that \lev\ is IR.
\end{proof}

\paragraph{Why \lev?}

The illustration of \lev{} serves two purposes: (a)~it shows that an IC network auction can have quite high diversity in its design, and (b)~this diversity can be exploited in order to earn higher revenue from an auction. In the experiments (\Cref{sec:experiment}), we show this via simulations. However, tuning the mechanism for a higher revenue requires prior information on the valuations.

} 

\section{Bayesian Setup and Optimal Auction}
\label{sec:optimal-trees}

The optimal auction is the one that maximizes the expected revenue. This is done assuming that the prior of the valuations are known to the auctioneer, which is a common assumption in classical auction literature~\citep[e.g.]{Myerson1981}.\footnote{This assumption is primarily due to two reasons: (a)~for prior-free auctions, the worst-case revenue can be arbitrarily bad, hence revenue maximization does not yield any useful result, and (b)~in practice, the prior on the users' valuation can be estimated from the historical data.} In this section, we consider the revenue-optimal auction where the prior distribution over $(v_i, v_{-i})$ is given by $P$ and is a common knowledge.
First, we define the notion of truthfulness in the prior-based setup.

\begin{definition}[Diffusion Bayesian Incentive Compatibility]
\label{def:DBIC}
 A \da{} $(g,p)$ on a graph $G$ is {\em diffusion Bayesian incentive compatible (\dbic)} if
 \begin{enumerate}
     \item every agent's expected utility is maximized by reporting her true valuation irrespective of the diffusing status of herself and the other agents, i.e., for every $i \in N$, $\forall r_i, \hat{r}_{-i}$, the following holds
     \begin{align*}
     \lefteqn{
     \mathbb{E}_{v_{-i}|v_i} \left[ v_i g_i(f^G((v_i,r_i'), (v_{-i},\hat{r}_{-i}))) - p_i(f^G((v_i,r_i'), (v_{-i},\hat{r}_{-i}))) \right] } \\
     &\geqslant \mathbb{E}_{v_{-i}|v_i} \left[ v_i g_i(f^G((v_i',r_i'), (v_{-i},\hat{r}_{-i}))) - p_i(f^G((v_i',r_i'), (v_{-i},\hat{r}_{-i}))) \right], \forall v_i, v_i', r_i' \subseteq r_i, \textup{ and,}
     \end{align*}
     \item for every true valuation, every agent's expected utility is maximized by diffusing to all its neighbors irrespective of the diffusion status of the other agents, i.e., for every $i \in N$, $\forall r_i, \hat{r}_{-i}$, the following holds
     \begin{align*}
         \lefteqn{ \mathbb{E}_{v_{-i}|v_i} \left[ v_i g_i(f^G((v_i,r_i), (v_{-i},\hat{r}_{-i}))) - p_i(f^G((v_i,r_i), (v_{-i},\hat{r}_{-i})))  \right] } \nonumber \\
         &\geqslant \mathbb{E}_{v_{-i}|v_i} \left[ v_i g_i(f^G((v_i,r_i'), (v_{-i},\hat{r}_{-i}))) - p_i(f^G((v_i,r_i'), (v_{-i},\hat{r}_{-i})))  \right], \forall v_i,r_i' \subseteq r_i.
     \end{align*}
 \end{enumerate}
\end{definition}

It is easy to see that \ddsic{} implies \dbic{} since \dbic{} requires Conditions~\ref{point1} and \ref{point2} of \Cref{def:DDSIC} to hold only in expectation.

\subsection{Characterization of \dbic\ Mechanisms}

Our first result is to characterize the \dbic\ auctions. For convenience, we define the {\em expected} allocation and payments with the shorthand notation as described below.
\begin{align}
     \alpha_i((v_i,r_i'),\hat{r}_{-i}))&=\mathbb{E}_{v_{-i}|v_i}[  g_i(f^G((v_i,r_i'), (v_{-i},\hat{r}_{-i}))) ] \hspace{20pt} \textup{\bf (allocation)} \label{eq:bic-alloc} \\
     \ppp_i((v_i,r_i'),\hat{r}_{-i}))&=\mathbb{E}_{v_{-i}|v_i}[  p_i(f^G((v_i,r_i'), (v_{-i},\hat{r}_{-i}))) ] \hspace{20pt} \textup{\bf (payment)} \label{eq:bic-payment}
\end{align}

In the Bayesian setup, the notion of participation guarantee is also weakened to {\em interim individual rationality (IIR)} where the expected utility of a player to join the mechanism is non-negative after she learns her own type.
\begin{definition}[Interim Individual Rationality]
\label{def:IIR}
 A \da\ $(g,p)$ on a graph $G$ is {\em interim individually rational (IIR)} if  $v_i \alpha_i((v_i,r_i),\hat{r}_{-i})) - \ppp_i((v_i,r_i),\hat{r}_{-i})) \geqslant 0, \ \forall v_i, \hat{\theta}_{-i}, r_i, \forall i \in N$, where $r_i$ is the true neighbor set of $i$.
\end{definition}
Similar to \dbic, it is easy to see that IR implies IIR since IIR requires the conditions of \Cref{def:IR} to hold only in expectation.
Next, we define the following structure of an auction to succinctly characterize the \dbic\ auctions.

\begin{definition}[Monotone and Forwarding-Friendliness in Expectation (\mffe{})]
 \label{def:ffep}
 For a given network $G$, a \da\ $(g,p)$ is {\em monotone and forwarding-friendly in expectation (\mffe)} if
 \begin{enumerate}[(a)]
    \item \label{cond1-dbic} for every $i \in N$ and $r_i,\hat{r}_{-i}$, the functions
     $\alpha_i((v_i,r_i'),\hat{r}_{-i})$ is non-decreasing in $v_i$, for every $r_i' \subseteq r_i$, and for the given allocation function $\alpha$, the payment $\ppp_i$ for each player $i \in N$ is such that, for every $v_i,r_i$, and $\hat{r}_{-i}$, the following two conditions hold.
     \item \label{ffm(b)-dbic} For every $r_i' \subseteq r_i$, the following payment formula is satisfied.
\begin{gather}
    \ppp_i((v_i,r_i'),\hat{r}_{-i}) =
       \ppp_i((0,r_i'),\hat{r}_{-i})+ v_i \alpha_i((v_i,r_i'),\hat{r}_{-i}) -\int_0^{v_i}\alpha_i((y,r_i'),\hat{r}_{-i}) \diff y \label{eq:dbic-int}
\end{gather}
     \item \label{ffm(a)-dbic} The values of $\ppp_i((0,r_i'),\hat{r}_{-i})$ and $\ppp_i((0,r_i),\hat{r}_{-i})$ are arbitrary real numbers that satisfies the following inequality for every $r_i' \subseteq r_i$.
\begin{gather}
\ppp_i((0,r_i'),\hat{r}_{-i})-\ppp_i((0,r_i),\hat{r}_{-i}) \geqslant \int_0^{v_i}(\alpha_i((y,r_i'),\hat{r}_{-i})-\alpha_i((y,r_i),\hat{r}_{-i})) \diff y\label{eq:constraint-dbic}
\end{gather}
 \end{enumerate}
\end{definition}

\begin{theorem}[\dbic\ Characterization]
\label{thm:characterization-dbic}
 A \da\ $(g,p)$ is \dbic{} if and only if it is \mffe{}.
\end{theorem}

\begin{psketch}
 The proof of the direction \mffe{} $\Rightarrow$ \dbic{} is identical to \Cref{thm:reverse_characterization} with the allocations and payments, $g_i$ and $p_i$, replaced with their expected versions, $\alpha_i$ and $\ppp_i$ (\Cref{eq:bic-alloc,eq:bic-payment}), respectively. The other direction is almost identical to \Cref{thm:characterization} since the same starting inequalities also hold for \dbic{} with the above-mentioned replacements, i.e., $g_i$ and $p_i$, replaced with their expected versions, $\alpha_i$ and $\ppp_i$ (\Cref{eq:bic-alloc,eq:bic-payment}), respectively.
 We skip rewriting the almost identical steps with the above-mentioned substitutions.
\end{psketch}

\subsection{\edit{Referral auctions}}
\label{sec:referral}

\edit{
Multi-level marketing (MLM) is a marketing approach that incentivizes individuals who not only adopt a product but advertise it also~\citep{Emek11,Drucker2012,Babaioff2012}. On a social network, it creates a viral effect where the information regarding a product reaches far beyond what traditional marketing can do. Due to its similarity with the objective of diffusion auctions, i.e., to spread the information of the auction to more individuals on a network, in this section, we consider a natural adaptation of MLM into auctions and call this class of auctions as \emph{referral auctions} (\ra{}).

In a referral auction, the seller invites its immediate neighbors in the network to report their valuations and invite all of their neighbors. These agents are also suggested, in turn, to spread the same message, i.e., to ask their neighbors to report their valuations and forward the information to their neighbors. Each time a node $i$ reports and forwards the information to its neighbors, the information of $(\hat{v}_i, \hat{r}_i)$ is recorded by the seller along with its (system-generated) timestamp $\tau_i$. Note that one can implement this auction in various possible ways, e.g., via inviting each node to register on the seller's site and providing the information of their neighbors. In all possible such cases, the seller can record the timestamp which cannot be manipulated by the agents. This information will be used by the class of referral auctions.

In the class \ra{}, all agents are sorted w.r.t.\ their timestamps and a referral tree is formed via a first-invite-first-served policy (breaking ties in a fixed order, e.g., w.r.t.\ their social IDs). This implies that the unique parent of every node is determined by the earliest timestamp of those inviting nodes. Note that, this is the principle of multi-level marketing as well -- only those individuals on a network are considered for referral bonuses that invited a new customer {\em first} to the seller's system.

Once the referral tree is formed, the mechanism runs an auction at every level of the tree through a general {\em deterministic} allocation rule $\gggg$ which is monotone non-decreasing and runs only on the agents at a given level. Define the corresponding payment as
\begin{equation}
    \label{eq:gen-lbl-payment}
    \pppp_i(\rho_i,\rho_{-i}) = \rho_i \gggg_i(\rho_i,\rho_{-i}) -\int_0^{\rho_i}\gggg_i(y,\rho_{-i}) \diff y.
\end{equation}
We note that the payment formula in \Cref{eq:gen-lbl-payment} is the same as the payment formula in the classical result of \citet{Myerson1981} with the \vipc\ term being zero.
Based on different choices of $\gggg$, we obtain the class \ra{}, described algorithmically in \Cref{alg:lblg}.

}

\begin{algorithm}[t]
\caption{Referral Auctions (\ra)}
\label{alg:lblg}
 \KwIn{reported types $\hat{\theta}_i = (\hat{v}_i, \hat{r}_i), \hat{r}_i \subseteq r_i$, \edit{and recorded timestamps $\tau_i$}, for all $i \in N$}
 \Parameter{\edit{an arbitrary monotone non-decreasing deterministic allocation $\gggg$}}
 \KwOut{winner of the auction (which can be $\emptyset$), payments of each agent}
 {\bf Preprocessing:} \edit{Create the referral tree $\hat{T}$ rooted at $s$ such that the neighbors of $s$ is $\ch(s)$, and $\pa(i) = \argmin \{\tau_k : i \in \hat{r}_k\}$, for all $i \in N \setminus \ch(s)$. Ties are broken w.r.t.\ a fixed order over the nodes.}\; \label{line-prep-g}
\uIf{$\hat{v}_i = 0, \forall \ i \in N$\label{line00-g}}
    {Item is not sold and payment is set to zero for all agents, STOP\;}
{\em Initialization:} all agents are \texttt{non-winners} and their \texttt{actual payments} are zeros, set \texttt{offset} = 0, \texttt{level} = 1, \texttt{parent} = $s$, $v_\pa = 0$\;
In this level of $\hat{T}$: \; \label{stepfirst-g}
\Indp
\For{each node $i \in \ch(\pa)$}{
    Set \texttt{effective valuation} $\rho_i := \max\{\hat{v}_j : j \in \hat{T}_i\} - \of$\;
}
Remove the nodes that have $\rho_i < 0$, denote the rest of the agents with $N_\textup{\tt remain}$\; \label{remove-g}
\uIf{$|N_\textup{\tt remain}| \geqslant 2$}{
Find $i^*$ where $\gggg_{i^*}(\rho_{i^*}, \rho_{N_\textup{\tt remain} \setminus \{i^*\}}) = 1$ \;
Compute $z := \pppp_{i^*}(\rho_{i^*}, \rho_{N_\textup{\tt remain} \setminus \{i^*\}})$, given by \Cref{eq:gen-lbl-payment}\;
}
\uElse{
Set $z=0$\;
} \label{line09-g}
\uIf{$v_{\pa}\geqslant \of +z$\label{line10-g}}
    {STOP and go to Step~\ref{finalstep-g} \;}
Set agent $i^*$ as the \texttt{tentative winner} and its \texttt{effective payment} to be $z$ \;\label{line13-g}
All nodes and their subtrees except $i^*$ are declared \texttt{non-winners} \;
The \texttt{actual payment} of $i^*$ to \texttt{parent} $=$ \texttt{effective payment} + \texttt{offset} \;
\texttt{parent} = $i^*$, $\texttt{offset} =  \text{ \texttt{actual payment} of } i^* $ \; \label{steppayment-g}
\texttt{level} = \texttt{level} + 1 \; \label{steplast-g}
\Indm
Repeat Steps~\ref{stepfirst-g} to \ref{steplast-g} with the updated \texttt{parent} and \texttt{offset} for the new level \;
STOP when no agent $i$ has $\rho_i \geqslant 0$ OR the leaf nodes are reached \; \label{line18-g}
Set \texttt{tentative winner} as \texttt{final winner}; final payments are the \texttt{actual payments} that are paid to the respective parents of $\hat{T}$ \; \label{finalstep-g}
\end{algorithm}

We show that each member of \ra{} also follows the desirable properties like \lev{}. Since the proof is quite similar to that of \lev, we provide the sketch to show exactly the places where the proof differs.
\edit{Note that the mechanisms in \ra{} generate a referral tree $\hat{T}$ from an arbitrary underlying network. Hence, to prove truthfulness of the auctions in this class, we need to show that no agent can profit by underreporting her set of {\em true} neighbors in the underlying graph.}

\edit{
\begin{theorem}
    \label{thm:misreport-ref-tree}
    In each auction in \ra{}, no agent $i \in N$ gets a higher utility by reporting $\hat{r}_i \subset r_i$.
\end{theorem}

\begin{proof}
    Each auction in \ra{} is designed in such a way that only the \opnw{} or the \win{} gets a non-negative utility. Each \nopnw{} node gets a utility of zero. Also, note that the auctions in \ra{} create the referral tree in a first-invite-first-served manner. Since the agents cannot alter their timestamps, if they under-report their neighbor set, they can potentially stop becoming a \opnw{}, which does not improve their utility. This observation is the key to this proof. Consider the following four cases for an agent $i$.

    \smallskip
    \emph{Case 1: agent $i$ is the \win{}.} In this case, her forwarding information is irrelevant to her utility. Hence, it does not violate the condition of the theorem.

    \smallskip
    \emph{Case 2: agent $i$ is a \nopnw{} after reporting $r_i$, her true neighbor set.} In this case as well, her forwarding information is irrelevant to her utility. This is because, even after reporting her entire neighbor set, she was not on the path to the winner. By misreporting, agent $i$ will continue to be a \nopnw{} and her utility will continue to be zero.

    \smallskip
    \emph{Case 3: agent $i$ is an \opnw{} when it reports $r_i$, but a \nopnw{} when it reports $r_i' \subset r_i$.} In this case, the utility is zero when agent $i$ is a \nopnw{}, but her utility is non-negative when she is an \opnw{}. Therefore, agent $i$ cannot improve her utility in this case as well.

    \smallskip
    \emph{Case 4: agent $i$ is an \opnw{} when it reports $r_i$, and continues to be a \opnw{} when it reports $r_i' \subset r_i$.} According to \Cref{alg:lblg}, agent $i$ gets the same utility in both these cases.

Hence, we conclude that in all possible cases of misreported neighbor set that alter the referral tree $\hat{T}$, an agent cannot obtain a better utility.
\end{proof}

}

\edit{
\begin{theorem}
 \label{thm:gen-lbl}
 Each auction in \ra{} is \ddsic{} and IR.
\end{theorem}
}

\begin{psketch}
 Consider an arbitrary auction $f \in \ra{}$. \edit{We showed in \Cref{thm:misreport-ref-tree} that an agent cannot manipulate the referral tree to her favor. Hence, we need to show that for the formed referral tree $\hat{T}$, $f$ satisfies \ddsic{} and IR.}
 The proof follows similar line of arguments as \Cref{thm:lblev-truthful,thm:lblev-ir} with a few variations, which we describe here. We follow the same definitions of \opnw{}, \nopnw{}, and \win{} for the different types of agents, and use the terms \of{}, \ch{}, and \pa{} as defined there. The \wi{} function is updated as $\wi(i) = \arg_{j \in \ch(i)} \{\gggg_{j}(\rho_{j}, \rho_{N_\textup{\tt remain} \setminus \{j\}}) = 1\}$, and there is no \ru{} function.

 \paragraph{Part 1: $f$ satisfies Condition~\ref{cond1} of \mff{} (\Cref{def:ffm}):}
 This part follows by the same arguments and the fact that $\gggg$ is monotone non-decreasing.

 \paragraph{Part 2 and 3: $f$ satisfies Conditions~\ref{ffm:a} and \ref{ffm:b} of \mff{} (\Cref{def:ffm}):}
 These two arguments ensure the payment formula and the condition on the \vipc{} terms. The same conditions can be obtained with the same set of arguments by replacing $\rho_\ell^{t_\ell / t_k}$ with $\inf \{\rho_i \in \mathbb{R} : \gggg_{i}(\rho_{i}, \rho_{N_\textup{\tt remain} \setminus \{i\}}) = 1\}$ at every level of the tree. Since $\rho_i$ is obtained by subtracting $\of(\pa(i))$ from agent $i$'s valuation and that agent is removed if this number is negative, $\rho_i$'s are non-negative by design. Hence, the number $\inf \{\rho_i \in \mathbb{R} : \gggg_{i}(\rho_{i}, \rho_{N_\textup{\tt remain} \setminus \{i\}}) = 1\}$ is also non-negative. Therefore, this number follows every argument that $\rho_\ell^{t_\ell / t_k}$ followed at every level of the proof of \Cref{thm:lblev-truthful} in an identical way.

 Collecting these three parts, we prove that $f$ is \ddsic.

 Similarly, the IR proof follows an identical set of arguments follow with the same substitution of $\rho_\ell^{t_\ell / t_k}$ with $\inf \{\rho_i \in \mathbb{R} : \gggg_{i}(\rho_{i}, \rho_{N_\textup{\tt remain} \setminus \{i\}}) = 1\}$ at every level of the tree. We skip rewriting the identical steps with the above-mentioned substitution.
\end{psketch}

Since each $f \in \ra{}$ is \ddsic\ and IR, they are \dbic\ and IIR. \edit{For simplicity of terminology, we will call each member of the class $\ra{}$ simply an \ra{} (referral auction) henceforth.}

\edit{
In what follows, we will find an \ra{} that maximizes the expected revenue of the seller when the valuations of the buyers are i.i.d. The high-level idea of our proof is the following. We observe from \Cref{eq:gen-lbl-payment} that the revenue of the seller in any auction in \ra{} is simply the sum of the payments made by the buyers at the first level, when their valuations are replaced with their effective valuations, i.e., the maximum valuation in their subtree.
This allows us to ``replace'' each buyer at the first level, with the buyer having a maximum valuation in its subtree. We formalize this idea in the proof of \Cref{thm:optimal-revenue-tree} and \Cref{lemma:reduction}.
}

\subsection{Optimal \edit{referral auction} for i.i.d.\ valuations}

In the objective of finding the {\em revenue-optimal} mechanism on a network, we address the problem in steps. In this section, we consider the mechanisms in the class \ra{}, assuming that the priors on the valuations are known to the designer and that all $v_i$'s are i.i.d.\ with distribution $F$ that follows the {\em monotone hazard rate} (MHR) condition, i.e., $f(x)/(1-F(x))$ is non-decreasing in $x$.\footnote{The intuitive meaning of this condition is that the distribution is not {\em heavy-tailed}. Many distributions, e.g., uniform and exponential, follow the MHR condition~\citep{barlow1963properties}.}
In this section, we find the optimal mechanism for this setup. To do that, first, we need to define a {\em transformed auction} (\ta) of an \ra{} as follows.
\edit{
\begin{definition}[Transformed Auction]
 A {\em transformed auction} (\ta) of an \ra{} is the auction where each subtree $\hat{T}_i, i \in \ch(s)$ is replaced with a node with a valuation of $\max_{j \in \hat{T}_i} v_j$, and the allocation and payments are given by $(\gggg_i,\pppp_i), i \in \ch(s)$.
\end{definition}
}
Note that a \ta\ does not specify the allocations beyond the first level of the tree. This is because, we will only be interested in the revenue generated by a \ta, and every \ta, regardless of how it allocates the object and extracts payments in the subsequent levels, will earn the same revenue, as shown formally in the following result.

\edit{
\begin{lemma}
 \label{lemma:reduction}
 The revenue earned by an \ra{} is identical to its \ta.
\end{lemma}

\begin{proof}
 Note that in an \ra{}, the net payment received by the seller $s$ comes directly from the nodes in the first level of the tree. The \of{} is zero, and the payment is calculated based on the maximum valuation in the subtree of the agents in the first level. The rest of the payments in the tree are internally adjusted within the nodes and does not reach the seller. Therefore, the total revenue earned by an \ra{} can be simulated by transforming every first-level nodes with their valuations replaced with the maximum valuation of their subtree and applying $(\gggg,\pppp)$ on those nodes. Hence, we have the lemma.
\end{proof}
}

Given the above lemma, we can, WLOG, look only at the \ta{}s for revenue maximization.
In the \ta{} of a given \ra{}, the revenue maximization problem is restricted to the first level of the tree. However, the nodes of this restricted tree are the {\em transformed nodes} whose valuations are the maximum valuations of their respective subtrees. For notational simplicity, we use a fresh index $\ell$ to denote these transformed nodes at the first level, i.e., for $\ch(s)$. The {\em transformed valuation} of $\ell$ is denoted by $v_\ell := \max_{j \in \hat{T}_\ell} v_j$. Again, to reduce notational complexity, the set of the players in this \ta{} is represented by $\Nfirst := \ch(s)$.
%
%
%
%
In the following, we state the fact that the $v_\ell$'s also follow the MHR property.

\begin{fact}
 The {\em maximum} of a finite number of i.i.d.\ random variables, each of which satisfies the MHR condition, also satisfies the MHR condition.
\end{fact}

\begin{proof}
 Suppose, there are $n$ i.i.d.\ random variables given by $X_1, X_2, \ldots, X_n$, and their distribution is given by $F$. Let $Y := \max \{ X_1, X_2, \ldots, X_n \}$. It is given that $F$ satisfies MHR condition, i.e., $f(x) / (1-F(x))$ is monotone non-decreasing. Denote the distribution and density of $Y$ by $\tilde{F}$ and $\tilde{f}$ respectively. Now,
 \begin{align*}
     \tilde{F}(x) &= P (Y \leqslant x) = P (\max \{ X_1, X_2, \ldots, X_n \} \leqslant x) \\
     &= P(\cap_{i = 1}^n \{X_i \leqslant x\}) = \prod_{i = 1}^n P( X_i \leqslant x) = F^n(x). \\
     \text{Hence, } \tilde{f}(x) &= nF^{n-1}(x) f(x). \\
     \text{Therefore, } \frac{\tilde{f}(x)}{1-\tilde{F}(x)} &= \frac{nF^{n-1}(x) f(x)}{1-F^n(x)} \\
     &= \frac{f(x)}{1-F(x)} \cdot \left( \frac{n}{1 + \frac{1}{F(x)} + \frac{1}{F^2(x)} + \ldots + \frac{1}{F^{n-1}(x)}}\right).
 \end{align*}
 Since $F(x)$ is non-decreasing, $1/F(x)$ is non-increasing. Hence, the denominator of the last term in the last expression is also non-increasing, leading the expression to be non-decreasing. Since $\frac{f(x)}{1-F(x)}$ is non-decreasing as well, we conclude that $\frac{\tilde{f}(x)}{1-\tilde{F}(x)}$ is also non-decreasing and hence $Y$ satisfies MHR.
\end{proof}

We now focus on the revenue maximization problem. Note that, $\gggg$ and $\pppp$ are particular choices of the allocations $g$ and $p$ respectively. Therefore, the expected allocations and payments are given by \Cref{eq:bic-alloc,eq:bic-payment} with $g$ and $p$ replaced with $\gggg$ and $\pppp$ respectively. In particular, the \vipc{} term in \Cref{eq:bic-payment} is zero for the nodes in the \ta{} since the payment $\pppp$ sets it to zero for the nodes in the first level of the class \ra{}.\footnote{The \vipc{} term needs to be non-positive for the auction to be IIR, and since our objective is to maximize revenue, it must be zero. This is ensured by $\pppp_i$.} Also, in the \ta{}, the \of\ is zero. Therefore, $\rho_\ell = v_\ell, \ \forall \ell \in \Nfirst$. The neighbor component of the types $r_\ell$ are no longer relevant since the mechanism is restricted to the first level in the \ta{}. Hence, we can reduce the arguments of $\ppp_\ell$ and $\alpha_\ell$ to only $v_\ell$ in \Cref{eq:bic-alloc,eq:bic-payment}. Since, the only variable parameter in the payment of the agents is the allocation function $\gggg$, the optimization problem for revenue maximization in the \ra{} class is given by
\begin{equation}
    \label{eq:rev-max}
    \begin{aligned}
        \max \qquad & \sum_{\ell \in \Nfirst} \int_{v_\ell = 0}^{b_\ell} \ppp_\ell (v_\ell) f_\ell(v_\ell) \diff v_\ell \\
        \text{s.t.} \qquad & \gggg \text{ is monotone non-decreasing and deterministic}
    \end{aligned}
\end{equation}
In the above equation, $f_\ell$ is the density of $v_\ell$, which is assumed to have a bounded support of $[0,b_\ell]$. We will denote the corresponding distribution with $F_\ell$. This optimization problem now reduces to the classic single item auction setting of \citet{Myerson1981}. Following that analysis, we find that the individual terms in the sum of the objective function of \Cref{eq:rev-max} can be written as follows
\begin{align*}
    \lefteqn{\int_{v_\ell = 0}^{b_\ell} \ppp_\ell (v_\ell) f_\ell(v_\ell) \diff v_\ell} \\
    &= \int_{0}^{b_\ell} w_\ell (v_\ell) \alpha_\ell(v_\ell) f_\ell(v_\ell) \diff v_\ell \\
    &= \int_{0}^{b_\ell} w_\ell (v_\ell) \left( \int_{v_{-\ell}} \gggg_\ell(v_\ell, v_{-\ell}) f_{-\ell}(v_{-\ell}) \diff v_{-\ell} \right) f_\ell(v_\ell) \diff v_\ell \\
    &= \int_v w_\ell (v_\ell) \gggg_\ell(v_\ell, v_{-\ell}) f(v) \diff v.
\end{align*}
The expression $w_\ell(x) := x - (1-F_\ell(x)) / f_\ell(x)$ is defined as the {\em virtual valuation} of agent $\ell$ and for completeness, the derivation of the first equality is provided in the appendix. The second equality holds after expanding $\alpha_\ell(v_\ell)$ from \Cref{eq:bic-alloc}. The last equality holds since the valuations are independent (but may not be identically distributed as the number of nodes in the subtree of $\ell$ can be different from that of $\ell'$), and $f$ denotes the joint probability density of $(v_\ell, v_{-\ell})$.

The objective function of \Cref{eq:rev-max} can therefore be written as
\[\int_v \left( \sum_{\ell \in \Nfirst} w_\ell (v_\ell) \gggg_\ell(v_\ell, v_{-\ell}) \right) f(v) \diff v. \]
The solution to the unconstrained version of the optimization problem given by \Cref{eq:rev-max} is rather simple.
\begin{equation}
\label{eq:optimal-mech}
 \begin{aligned}
  \text{if } & w_\ell (v_\ell) < 0, \forall \ell \in \Nfirst, \text{ then } \gggg_\ell(v_\ell, v_{-\ell}) = 0,  \forall \ell \in \Nfirst \\
  \text{else } & \gggg_\ell(v_\ell, v_{-\ell}) = \left \{ \begin{array}{ll}
                                            1 & \text{ if } w_\ell (v_\ell) \geqslant w_k (v_k), \forall k \in \Nfirst \\
                                            0 & \text{ otherwise}
                                           \end{array}
                                    \right .
 \end{aligned}
\end{equation}
The ties in $w_\ell (v_\ell)$ are broken arbitrarily.
Since the distributions of $v_\ell, \ell \in \Nfirst$ satisfy MHR, the virtual valuations, $w_\ell$, are monotone non-decreasing. Also, since this mechanism breaks the tie arbitrarily in favor of an agent, the allocation is also deterministic. Therefore, the optimal solution of the unconstrained problem of \Cref{eq:rev-max} also happens to be the optimal solution of the constrained problem. We find the payments of the winner from \Cref{eq:gen-lbl-payment} as follows.
\begin{equation}
 \label{eq:optimal-payment}
 \begin{aligned}
    \text{define } \kappa_\ell^*(v_{-\ell}) &= \inf \{y : \gggg_\ell(y, v_{-\ell}) = 1\}, \\
    \pppp_\ell(v_\ell, v_{-\ell}) &= \kappa_\ell^*(v_{-\ell}) \cdot \gggg_\ell(v_\ell, v_{-\ell}),
 \end{aligned}
\end{equation}
where $\kappa_\ell^*(v_{-\ell})$ is the minimum valuation of agent $\ell$ to become the winner.
Formally, we define the auction as follows.
\begin{definition}[Maximum Virtual Valuation Auction (\mvv{})]
\label{def:mvv}
 The {\em maximum virtual valuation} auction is a subclass of \ra{}, where the \ta{}s of that subclass follow the allocation and payments given by \Cref{eq:optimal-mech,eq:optimal-payment} respectively.
\end{definition}
We consolidate the arguments above in the form of the following theorem.
\begin{theorem}
\label{thm:optimal-revenue-tree}
 For agents having i.i.d.\ MHR valuations, the revenue-optimal \ra{} is \mvv{}.
\end{theorem}

\edit{Since multiple \ra{}s can reduce to the same \ta{}, the revenue optimal \ra{} is a class of auctions, all belonging to \ra{}, that has the same \ta{} given by \Cref{def:mvv}.}
Note that neither IDM nor \lev{} is \mvv{} because they do not use any priors. Therefore, the revenue-maximizing auctions in this setting are a new class of mechanisms that have not been explored in the literature.



\subsection{Extension to non-i.i.d.\ agents and any diffusion auction}
\label{sec:extension}

When we migrate away from i.i.d.\ valuations, it is not clear if the nodes in the \ta{} satisfy MHR or a relatively weaker condition of {\em regularity} (which only requires the virtual valuations to be non-decreasing). Hence, the revenue maximization problem becomes far more challenging.

We provide an experimental study in the next section that shows that if the i.i.d.\ assumption does not hold, a special auction from the \lev{} class can yield more revenue than the currently known network auctions.

\edit{
To generalize our results beyond \ra{} for revenue maximization, we need to consider the revenue maximization problem~\Cref{eq:rev-max} with the constraints of \mffe{}~(\Cref{def:ffep}). This optimization problem seems to have much less structure than that in \ra{}. Therefore, we need more structural results about the $\ppp_i$ terms so that this optimization problem can be simplified. Also, we cannot work with only \ta{}s anymore since the revenue-optimal \da{} may not be reducible to a \ta{} ({\em \`{a} la} \Cref{lemma:reduction}).
We address these problems in our future work.
}

\section{Designing \lev\ for Improving Revenue}
\label{sec:experiment}

In the previous section, we investigated the case of i.i.d.\ valuations. In this section, we show that if the valuations are non-i.i.d., \edit{there exists non-trivial diffusion auctions that yield higher expected revenue than the known diffusion auctions in literature.}

\lev\ generates a class of mechanisms (\Cref{alg:lblev}) for different choices of the exponents $t \in \mathbb{R}_{>0}^n$ and trees $\hat{T}$. While each mechanism in that class is \ddsic\ and IR, we can anticipate that certain choices may lead to a higher revenue collected by the auction. In this section, we test that hypothesis and find out how the exponents $t$ can be designed such that it improves the revenue over the other \ddsic\ and IR auctions known in the literature. The revenues are compared for the same tree $\hat{T}$ for all competing auctions.

Revenue maximization in an auction is typically done in a prior-based approach~\citep{Myerson1981,manelli2007multidimensional,hart2017approximate}. Such approaches assume that the distribution of the types is known by the mechanism designer from its past interactions with the agents. We adopt a similar prior-based approach. In the context of network auctions, the types of the agents consist of the valuations and their neighbors' set and we assume a prior on both.

\noindent
\textbf{Other \ddsic\ mechanisms to compare.}
\citet{yuhangguodonghao2021} provide a comprehensive survey of the auctions on the network. TNM and CDM reduce to IDM when restricted to a tree. The objective of FDM and NRM is to redistribute the revenue so that there is little surplus -- hence, they are unsuitable for a revenue comparison. Other mechanisms like MLDM, CSM, and WDM apply to very specific settings, e.g., distribution markets with intermediatries, economic networks, and weighted networks respectively. Hence, we find that IDM is the sole candidate to compare with \lev.

\noindent
\textbf{Two-stage tree generation model.}
For the experiments, we need to iterate over randomly generated trees connecting the agents, and also equip the mechanism designer with some prior information about the connections model. To do so, we adopt a two-stage tree generation model, where the first stage is observable by the designer, but the second stage is not. It resembles the situation: {\em the designer knows who can probably be the children of which nodes, but cannot deterministically observe it while designing the mechanism}.

The mechanism \lev\ extracts a tree $\hat{T}$ from the graph generated through the reported $r_i$'s. For our experiments, we create the $r_i$'s in two stages. First, we generate a {\em base} tree in the following way.
We fix the number of nodes $n$ in this tree. Starting from the seller, which is not a part of the agent set, a random {\em children set} is picked for each node at every level of the tree where the size of the children set is drawn uniformly at random between $1$ and $\lfloor n/3 \rfloor$ from the rest of agents without replacement. This process is continued until all the $n$ nodes are exhausted (the last parent node gets the remaining number of nodes when it draws more than that remaining number).

The second stage probabilistically {\em activates} the edges on this base tree. The activated set of edges provides a sub-tree of the base tree, and this is considered to be the final tree $\hat{T}$.
More concretely, once the base tree is realized, the actual set $r_i$ is generated by tossing a coin with probability drawn from $\textup{Beta}(5,1)$ for each edge from $i$ to its children.
We assume that the second stage of randomization is not part of the prior information available to the mechanism designer, and hence, the choice of the exponent vector $t$ cannot depend on it. The second stage, therefore, helps us to cross-validate the mechanism designed from the prior information.

\noindent
\textbf{Valuation generation model.}
The valuations are drawn independently from ${\cal N}(\mu, \sigma^2)$. We assume that there are {\em three} classes of agents: \texttt{high}, \texttt{medium}, and \texttt{low}, having $\mu$ to be $100$, $70$, and $50$ respectively, and the same $\sigma$.
We will see the effect of $\sigma$ on the revenue in our experiments.

The prior information for the designer consists of the first stage of the tree generation process and the valuation generation process.
%

\noindent
\textbf{Setting the exponent vector $t$.}
Suppose, we knew that the nodes $i^*$ and $\ell$ are the first level nodes whose subtrees are the \wi\ and \ru\ respectively in \lev.
From \Cref{alg:lblev}, we know that the effective valuations $\rho_i$ of a first level node $i$ in the tree $\hat{T}$ is the maximum valuation of the nodes in the subtree of $i$, i.e., $\hat{T}_i$.
Note that, the revenue generated in \lev\ is $\rho_\ell^{t_\ell / t_{i^*}}$. Hence, a larger exponent ratio $t_\ell / t_{i^*}$ yields a better revenue as long as $\rho_\ell^{t_\ell / t_{i^*}} \leqslant \rho_{i^*}$. This is the driving philosophy of the following choice of $t$ with only the prior information.

On the base tree, we replace the nodes' valuations with the means, which is a prior information, and call the node having the highest and second highest $\rho_i$'s in the first level of the base tree to be the {\em expected} \wi\ and \ru\ respectively. Suppose, these two mean valuations are $w_\wi^e$ and $w_\ru^e$ respectively. We set the exponent of the first level {\em expected} \ru\ is set to $(1-\lambda) \cdot 1 + \lambda \cdot \frac{\log w_\wi^e}{\log w_\ru^e}$, with $\lambda$ being a parameter chosen by the mechanism. This is a convex combination between $1$, which is the exponent ratio for IDM, and the other extreme $\frac{\log w_\wi^e}{\log w_\ru^e}$. If the true \wi\ and the \ru\ would have indeed come from the subtrees of the expected \wi\ and \ru, then the exponent ratio $\frac{\log v_\wi}{\log v_\ru}$ would have extracted the maximum revenue in \lev. This is the intuition of using this factor as a candidate for the expected \ru's exponent.

The exponents of all other agents (including the expected \wi) are set to $1$.
Note that $t$ is decided based on the first stage of the tree generation process and the prior of the valuation. It is independent of the second level of the tree generation process, and therefore, is agnostic of the actual tree $\hat{T}$. Such a $t$ is indepedent of the agents' actions and is consistent with \Cref{alg:lblev}. Indeed there is a possibility that a probabilistic draw of the second stage of the tree generation process may have a different \wi\ and \ru\ than their expected ones, which makes this choice of $t$ sub-optimal than IDM for revenue.

\begin{figure}[t!]
    \centering
    \includegraphics[width=\linewidth]{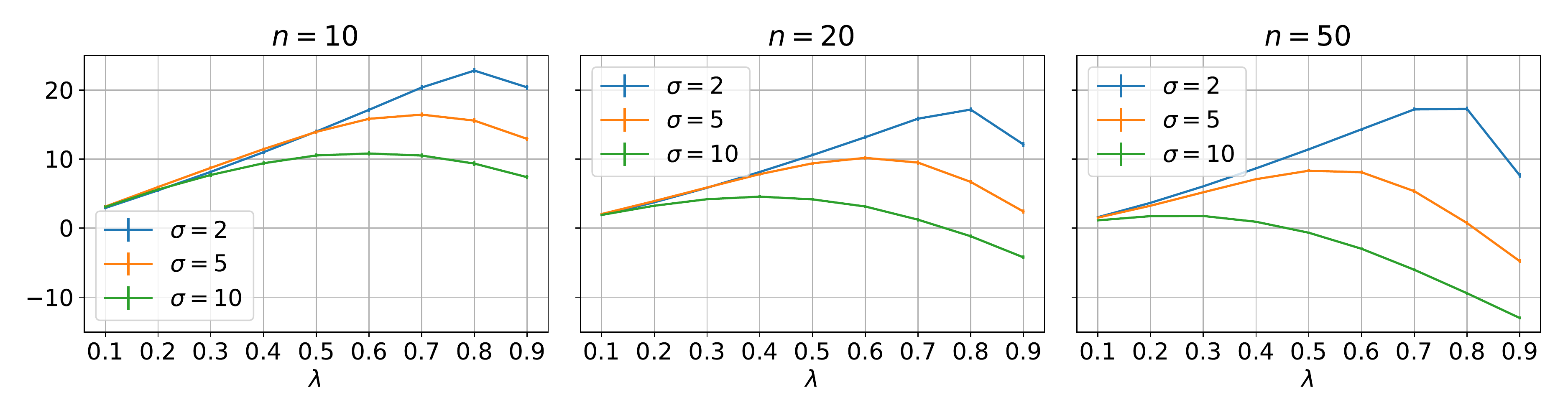}
    \caption{Percentage increase in revenue for \lev\ over IDM w.r.t.\ $\lambda$ for different $\sigma$ and $n$.}
    \label{fig:revenue_vs_lambda}
\end{figure}

\noindent
\textbf{First set of experiments.}
In this set of experiments, we find the effect of the three parameters, $n$, $\sigma$, and $\lambda$, on the revenue of the two auctions: (1)~\lev\ with the chosen exponent vector $t$ as above and (2)~IDM.
We consider one agent from class {\tt high}, $\lfloor (n-1)/2 \rfloor$ agents from class {\tt medium}, and $\lceil (n-1)/2 \rceil$ agents from class {\tt low}. This choice is to observe how the exponents $t$ make a difference in the revenue earned. If there are many agents of class {\tt high}, then it is highly probable that both the expected \wi\ and the \ru\ in the first level of the base tree has the same mean valuation, which makes the optimal exponent to be unity -- same as IDM.
In the first experiment, we consider different values for $\sigma$ and $n$, and compare the revenue earned by \lev\ and IDM with varying $\lambda$. The results are shown in \Cref{fig:revenue_vs_lambda}. For every $\lambda$, the base graph and the edge-activation probability (drawn from Beta(5,1)) generation have been repeated 100 times, and for each of such instances the edge activation and valuation generation for all agents have been repeated 100 times. The plot shows the mean percentage improvement of the revenue of \lev\ over IDM, with the standard error around it. Observe that, for every pair of $\sigma$ and $n$, there is an optimal convex combination (say $\lambda^*$) for which the revenue gap between \lev\ and IDM reaches a maxima.
Recall that this $\lambda^*$ also determines the exponent of the expected \ru\ in the base tree.

\begin{figure}[t!]
    \centering
    \includegraphics[width=0.5\linewidth]{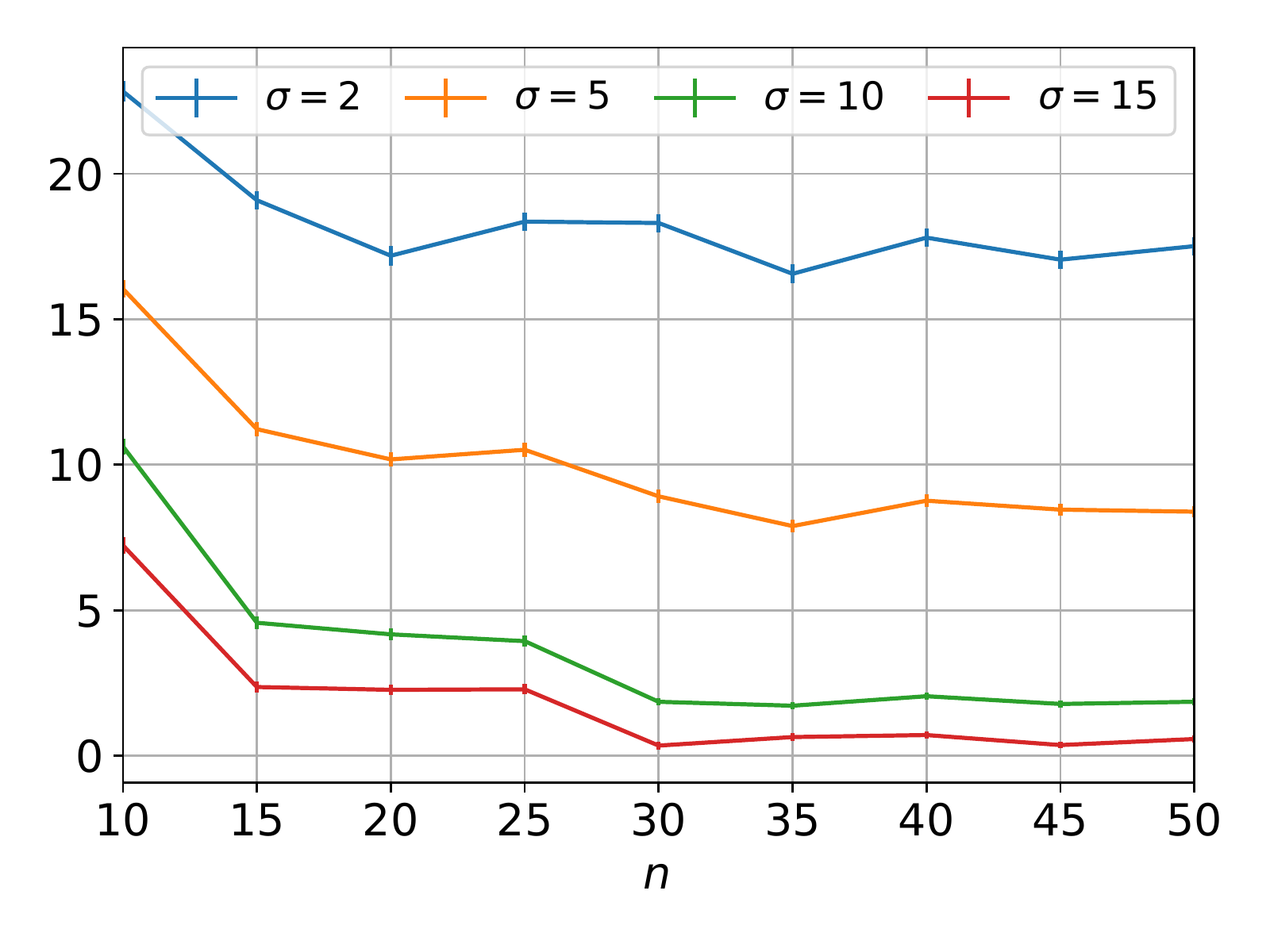}
    \caption{Percentage increase in revenue for \lev\ over IDM for different $\sigma$ and {\em learned} $\lambda^*$ (and hence $t$).}
    \label{fig:rev_vs_n_learned}
\end{figure}
\noindent
\textbf{Second set of experiments.}
The first set of experiments gives us the insight that the optimal convex combination factor $\lambda^*$ depends on $n$ and $\sigma$. Motivated by this observation, in this set of experiments, we run a regression model to learn the optimal $\lambda^*(n,\sigma)$ from several such instances of $(n, \sigma, \lambda^*)$. We used the {\em random forest regressor} \citep{breiman2001random} with the parameter of number of decision trees set to 100 to learn the $\lambda^*$ function.
We chose the random forest regressor for two reasons: (a)~from the examples, the function seems non-linear and instead of choosing a fixed non-linear function, an ensemble regressor could perform better, and (b)~random forest gave the best performance among the few other ensemble regressors we tested with (e.g., {\tt ADABOOST}, {\tt GradientBoosting}). We find that with the learned $\lambda^*$, which yields the exponents $t$, \lev\ performs better than IDM. For certain choices of $(\sigma, n)$, particularly when both $n$ and $\sigma$ are large, the learned exponents are close to $1$ for all agents, almost reducing \lev\ to IDM. \Cref{fig:rev_vs_n_learned} shows the results.

\section{Summary and Plans of Extension}
\label{sec:summary}

In this paper, we provided a characterization of randomized truthful single indivisible item auctions on a network. Our results are the network counterpart of Myerson's result \citep{Myerson1981}. We obtained a detailed description of the revenue optimal mechanism for a class called referral auctions with i.i.d.\ MHR valuations. When i.i.d.\ assumption does not hold, we provided a mechanism from our characterized class of \ddsic{} and IR auctions to experimentally show an improvement in the revenue from the currently known diffusion mechanisms. The question of finding the revenue optimal mechanism for a general network is still open and we want to pursue that as a future work.

%
%
%
\bibliographystyle{plainnat}
\bibliography{abb,swaprava,ultimate,references,master}


%
\appendix
\section*{Appendix}

\section{Derivation of the virtual valuation}

We need to show that
\begin{equation}
 \int_{0}^{b_\ell} \ppp_\ell (v_\ell) f_\ell(v_\ell) \diff v_\ell = \int_{0}^{b_\ell} w_\ell (v_\ell) \alpha_\ell(v_\ell) f_\ell(v_\ell) \diff v_\ell, \label{eq:integral-virtual-valuation}
\end{equation}
where,
\begin{align}
 \ppp_\ell (v_\ell) &= v_\ell \alpha_\ell(v_\ell) - \int_0^{v_\ell}\alpha_\ell(y) \diff y, \text{ and,} \label{eq:part-payment} \\
 w_\ell(v_\ell) &= v_\ell - \frac{1-F_\ell(v_\ell)}{f_\ell(v_\ell)}. \label{eq:part-virtual-valuation}
\end{align}
Substituting \Cref{eq:part-payment} in the LHS of \Cref{eq:integral-virtual-valuation}, we get
\begin{align*}
 \lefteqn{\int_{0}^{b_\ell} \ppp_\ell (v_\ell) f_\ell(v_\ell) \diff v_\ell} \\
 &= \int_{0}^{b_\ell} v_\ell \alpha_\ell(v_\ell) f_\ell(v_\ell) \diff v_\ell - \int_{0}^{b_\ell} \int_0^{v_\ell}\alpha_\ell(y) f_\ell(v_\ell) \diff y  \diff v_\ell \\
 &= \int_{0}^{b_\ell} v_\ell \alpha_\ell(v_\ell) f_\ell(v_\ell) \diff v_\ell - \int_{0}^{b_\ell} \int_y^{b_\ell}\alpha_\ell(y) f_\ell(v_\ell) \diff v_\ell \diff y \\
 &= \int_{0}^{b_\ell} v_\ell \alpha_\ell(v_\ell) f_\ell(v_\ell) \diff v_\ell - \int_{0}^{b_\ell} \alpha_\ell(y) \left( \int_y^{b_\ell}f_\ell(v_\ell) \diff v_\ell \right) \diff y \\
 &= \int_{0}^{b_\ell} v_\ell \alpha_\ell(v_\ell) f_\ell(v_\ell) \diff v_\ell - \int_{0}^{b_\ell} \alpha_\ell(y) (1 - F_\ell(y)) \diff y \\
 &= \int_{0}^{b_\ell} \left( v_\ell - \frac{1-F_\ell(v_\ell)}{f_\ell(v_\ell)} \right) \alpha_\ell(v_\ell) f_\ell(v_\ell) \diff v_\ell.
\end{align*}
In the second equality, we interchange the order of integration. The integrable space has the following order: $y$ varying from $0 \to v_\ell$ and thereafter $v_\ell$ varying from $0 \to b_\ell$ which is equivalent to $v_\ell$ varying from $y \to b_\ell$ and thereafter $y$ varying from $0 \to b_\ell$. The third equality holds by taking the $\alpha_\ell(y)$ term outside the inner integral since it is independent of $v_\ell$. The next equality holds since the distribution of $v_\ell$ has the support of $[0,b_\ell]$, hence at $b_\ell$, the value of $F_\ell$ is unity. The last equality is obtained by using the same integration varible for both integrals and rearranging them.

\end{document}